\documentclass[final,5p,times,twocolumn]{elsarticle}

\usepackage{graphicx,subfig,caption}

\usepackage{hyperref}

\hypersetup{colorlinks}

\usepackage[all]{hypcap}  

\usepackage{amssymb}

\usepackage{amsthm}

\usepackage{amsmath}  

\usepackage{lineno}

\usepackage{siunitx}

\journal{    }

\begin{document}

\begin{frontmatter}




\title{GEM-based TPC with CCD Imaging for Directional Dark Matter Detection}


\author{N. S. Phan \corref{cor1}}

\author{R. J. Lauer\corref{cor2}}
\author{E. R. Lee\corref{cor2}}
\author{D. Loomba\corref{cor2}}
\author{J. A. J. Matthews\corref{cor2}}
\author{E. H. Miller\corref{cor2} }

\address{Department of Physics and Astronomy, University of New Mexico, NM 87131, USA}
\cortext[cor1]{Corresponding author}

\begin{abstract}

The most mature directional dark matter experiments at present all utilize low-pressure gas Time Projection Chamber (TPC) technologies.  We discuss some of the challenges for this technology, for which balancing the goal of achieving the best sensitivity with that of cost effective scale-up requires optimization over a large parameter space.  Critical for this are the precision measurements of the fundamental properties of both electron and nuclear recoil tracks down to the lowest detectable energies.  Such measurements are necessary to provide a benchmark for background discrimination and directional sensitivity that could be used for future optimization studies for directional dark matter experiments.  In this paper we describe a small, high resolution, high signal-to-noise GEM-based TPC with a 2D CCD readout designed for this goal.  The performance of the detector was characterized using alpha particles, X-rays, gamma-rays, and neutrons, enabling detailed measurements of electron and nuclear recoil tracks.  Stable effective gas gains of greater than $1\times10^5$ were obtained in 100 Torr of pure CF$_{4}$ by a cascade of three standard CERN GEMs each with a 140 \si{\um} pitch.  The high signal-to-noise and sub-millimeter spatial resolution of the GEM amplification and CCD readout, together with low diffusion, allow for excellent background discrimination between electron and nuclear recoils down below $\sim$10 keVee ($\sim$23 keVr fluorine recoil).  Even lower thresholds, necessary for the detection of low mass WIMPs for example, might be achieved by lowering the pressure and utilizing full 3D track reconstruction.  These and other paths for improvements are discussed, as are possible fundamental limitations imposed by the physics of energy loss.

\end{abstract}

\begin{keyword}
dark matter \sep WIMPs \sep TPC \sep GEMs \sep $\text{CF}_{4}$ \sep CCD \sep directionality \sep head-tail effect \sep nuclear recoil \sep electron recoil
\end{keyword}

\end{frontmatter}


\section{Introduction}
\label{Introduction}

The nature of dark matter remains one of the most important unresolved questions in physics.  One of the leading candidates is a class of particles known as weakly interacting massive particles (WIMPs) \cite{griest, jungman}.  They are the target of many ongoing direct detection experiments, which aim to measure the signals left by the elastic scattering of WIMPs with nuclei in the detector target material \cite{goodman}.  A review of the many experimental searches for dark matter is found in Ref. \cite{gaitskell}.  Direct detection is limited by several factors, one being the extremely low WIMP-nucleon interaction cross-sections predicted by extensions of the Standard Model \cite{baer}, leading to a requirement of large detector masses.  Another is the featureless, exponentially-falling recoil energy spectrum expected from WIMP interactions, which encourages low detection thresholds.  These issues are compounded by the presence of backgrounds whose signals could mimic those expected from WIMPs.  Although powerful techniques have been developed to discriminate and shield against a majority of these backgrounds, the misidentification of backgrounds for signal continues to plague the field \cite{CRESST-II-UP, Kuzniak, Davis2014, Davis2015}.  For these reasons the definitive proof of discovery in dark matter searches rests on the detection of specific signatures of the WIMP-nucleus interaction arising from the Galactic origin of the WIMPs. 

One such textbook signature is the annual modulation in the interaction rate caused by the seasonal variation in the relative velocity of the Earth-bound detector with the dark matter halo \cite{drukier, spergel}.  The effect is relatively small (a few $\%$), however, and many known backgrounds also modulate seasonally \cite{Blum, Ralston, Borexino, FM, Davis}.  Although several experiments have observed an excess above expected backgrounds \cite{DAMA2008, DAMA-LIBRA-2013, CRESST-II-2012, CoGENT2013, CDMS-II-2013}, the results are inconsistent with null results obtained by others \cite{XENON100, SuperCDMS2014, LUX2016}.

A more robust and definitive Galactic signature is the sidereal modulation in the direction of the WIMP flux \cite{spergel}.  Due to the solar system's motion around the Galaxy, the flux appears to come from the direction of the constellation Cygnus, but as the Earth rotates through a sidereal day, the position of Cygnus, and hence, the direction of the incoming flux changes in the detector's frame of reference.  This signature is detected as a modulation in the mean nuclear recoil track direction, which is peaked in the direction opposite to Cygnus.  Not only is the directional signature more definitive in separating a dark matter signal from backgrounds, it is also much larger than the annual modulation effect due to the strong angular dependence of the nuclear recoil direction \cite{spergel}.  There are currently several underground experiments that have varying degrees of sensitivity to this directional signature, including DRIFT \cite{drift2000, drift2004}, NEWAGE \cite{miuchi}, MIMAC \cite{santos}, and DMTPC \cite{dujmic}.  In addition, there are a number of efforts performing R\&D  on directionality \cite{D3, emulsions2013, emulsions2014, nygrencolumnar, gehman, anisotropicscint2013}.  For a thorough review of directional dark matter detection see Refs. \cite{battatwhitepaper} and \cite{mayet}.

\section{Directional Detection Challenges}
\label{Directional Detection Challenges}

The main challenge for directional detection is that the low energy, 10's of keV (henceforth, keVr), WIMP-induced nuclear recoil tracks are extremely short in liquids and solids (10's - 100's \si{\nm}).  Thus, although R\&D is underway to develop technologies for solid \cite{emulsions2013, emulsions2014, anisotropicscint2013} and high pressure gas targets, \cite{nygrencolumnar, gehman}, most experiments use low pressure gas targets where directionality has been demonstrated \cite{driftheadtail, dmtpcheadtail}.  In this low energy regime, the recoiling nucleus will produce only a few hundred to a few thousand ionization pairs in the detection medium, with track lengths on the order of a millimeter even at pressures below 100 Torr (0.13 atm).  Consequently, a natural choice for technology, which is currently employed by all gas-based directional experiments, has been the Time Projection Chamber (TPC) invented by Nygren \cite{nygrentpc}.  This allows for full 3-dimensional (3D) reconstruction of the recoil track, together with the flexibility to choose gas targets and operating pressures over a broad range.  With cubic meter scale TPCs, the DRIFT experiment has pioneered the use of this technology for directional searches. With detectors that have a demonstrated directionality signature down to recoil energies of $\sim$40 keVr \cite{driftheadtail, drift2009}, DRIFT has set competitive limits on spin-dependent interactions for dark matter \cite{driftlimits2014}.

Nevertheless, many challenges remain, not least of which is the scalability of the current generation of directional experiments to reach the sensitivity required to test future claims of detection by non-directional experiments.  To accomplish this, an emphasis on maximizing sensitivity will need balancing with the scalability, cost, and robustness of the technology.  In this work we focus on measuring the intrinsic properties of low energy electron and nuclear recoil tracks and how they place fundamental limitations on the sensitivity of directional dark matter searches.  The nature of these tracks determine energy thresholds for both discrimination and directionality, with the latter defined as the energy at which the directional signature becomes detectable.  As we show in this work, these two energy thresholds are not the same, with the onset of directionality having a higher energy threshold than discrimination.  It is important to keep this in mind because, all else being equal, the true determinant of sensitivity for directional dark matter searches is the directionality threshold.  For the gas-based TPC detectors of interest here, the physical processes that affect the overall sensitivity - for both discrimination and directionality - involve energy loss, straggling, diffusion, signal generation (gas gain), and readout resolution.  These are briefly discussed here.

The direction of a recoiling atom in a gas-based TPC is reconstructed from the ionization track produced along its path.  For both electrons and nuclear recoils, this track is never straight due to multiple scattering (or straggling) with the constituents of the gas.  This results in a loss of resolution which varies with gas pressure, the energies and the masses of the recoiling particle and gas atoms.  For example, a 40 keVr fluorine recoil in 100 Torr CF$_4$ has an average range of 400 \si{\um} but suffers significant range straggling, defined as the standard deviation of the particle's final position relative to its initial direction, of about 150 \si{\um} and 110 \si{\um} in the longitudinal and lateral directions, respectively \cite{ziegler}.  

Following the generation of the ionization track, the negative charge drifts in a uniform electric field down to the TPC readout plane, where it undergoes amplification before being read out by the data acquisition.  During the drift, the ionization that defines the track undergoes additional loss of resolution due to diffusion, which depends on the drift distance, the strength of the drift field, and the nature of the drifting charge carrier and gas.  With electron drift, a select few gases such as carbon tetrafluoride (CF$_4$) have relatively low diffusion, which, for a drift distance of 20 cm, an electric field of 400 V/cm, and a pressure of 100 Torr, is approximately $\sigma_{\text{T}} = 790$ \si{\um} (recommended by Ref. \cite{Christophorou1} based on a fit to the measurements of Refs. \cite{Curtis, Naidu, Lakshminarasimha}) and $\sigma_{\text{L}} = 580$ \si{\um} (values in Ref. \cite{Christophorou2} derived from measurements in Ref. \cite{Hayashi}) for the transverse and longitudinal dimensions, respectively.

Even lower diffusion is possible for negative ion drifting gases such as carbon disulfide (CS$_2$), whose molecules have a high electron affinity.  These molecules capture the primary electrons to form negative ions that drift with a very low velocity and low diffusion due to thermal coupling with the bulk gas.  The use of an electronegative gas to suppress diffusion without a magnetic field was first proposed by Martoff et al. \cite{Martoff2000}, and measurements of mobility and diffusion in CS$_2$ mixtures \cite{Martoff2000, Ohnuki2001, Martoff2005, Pushkin2009, Dion2011, SI2013} indeed indicate thermal behavior for drift fields up to where measurements exist. Using the diffusion temperatures reported in Ref. \cite{SI2013} for CS$_2$,  the transverse and longitudinal diffusion widths are 320 \si{\um} and 330 \si{\um}, respectively, for a drift length of 20 cm at a field of 1000 V/cm.  Besides this clear advantage from a resolution standpoint, detector operation in the presence of high voltages and fields is quite stable in low pressure CS$_{2}$ (electrical discharges are suppressed).  However, one of the downsides of CS$_{2}$ is its lack of spin targets (if a spin dependent search is the goal), which necessitates mixing it with a gas such as CF$_{4}$ that contains $^{19}$F, a target with high nuclear spin content. 

Finally, after the charge arrives at the readout plane it must undergo avalanche amplification before being read out.  Here the large gas gain, needed for high signal-to-noise, is not easily achieved due to electrical instability at the low gas pressures required for directionality.  For example, in Multi-Wire Proportional Chambers (MWPCs) used in the DRIFT detectors, typical gas gains are $\lesssim$ 1000 \cite{battatJINST2014}.  But much higher gas gains have been achieved in other avalanche devices at low pressures \cite{shalemTHGEM1, shalemTHGEM2}.  Thus, the dependence of the gas gain on the avalanche device is an important factor to consider when designing a high signal-to-noise detector, particularly in the low pressure regime. 

\begin{figure*}[]
	\centering
	\subfloat[CCD detector]{\includegraphics[width=0.45\textwidth]{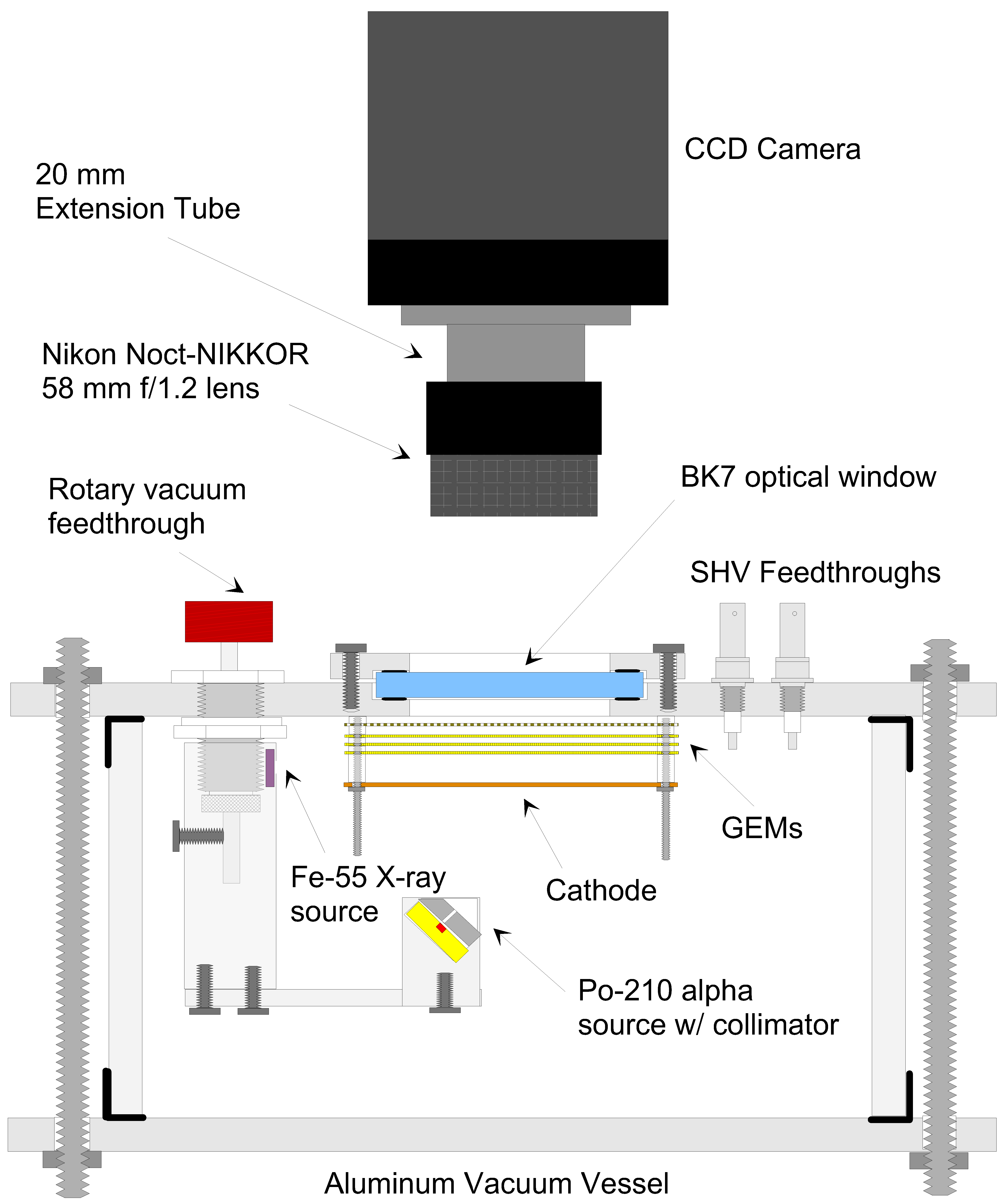}
	\label{fig:detectorGlobal}}
	\qquad
	\subfloat[Detection region]{\includegraphics[width=0.45\textwidth]{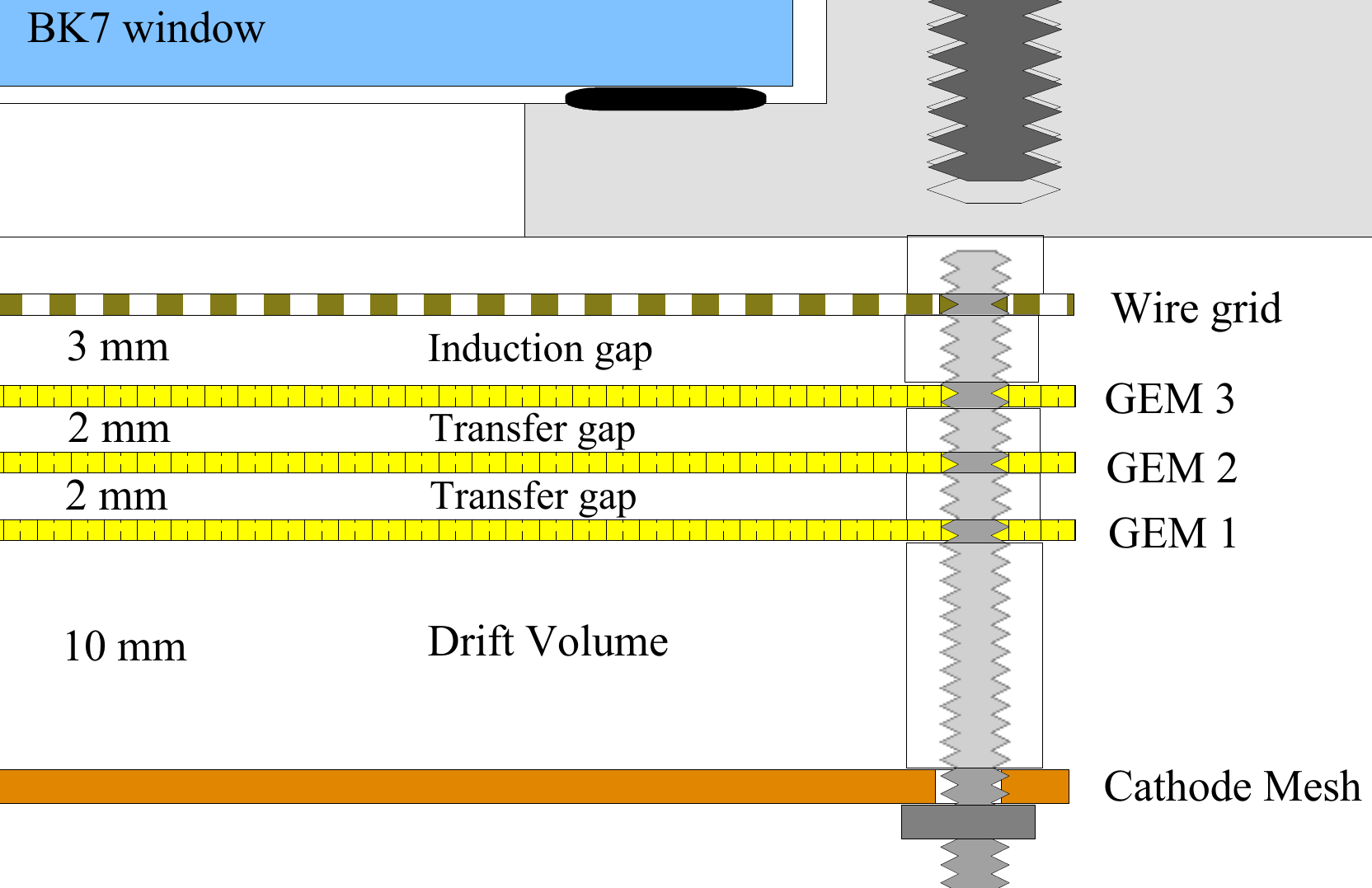}
	\label{fig:detectorGEM}}
	\caption[]{(a) A schematic of the CCD detector showing the relative positions of detector components.  The optical system, which consists of the CCD and lens, sits outside the vacuum vessel and images only the central 2.8 cm $\times $ 2.8 cm region of the topmost GEM surface.  The internally mounted $^{55}$Fe and $^{210}$Po sources are used for calibrations.  (b) A close up view of the detection volume and GEM stack showing the dimensions of the important regions.}
	\label{fig:detector}
\end{figure*}

Most physical processes, like those described above, degrade the directional sensitivity.  There is one, however, that has the potential to greatly enhance sensitivity if an intrinsic asymmetry in the energy loss of the recoil exists and is detectable.  Such an asymmetry would provide a means for assigning a head-tail (HT) sense to the track. The implication for directional sensitivity, based on simulations, is that with accurate head-tail, or vector, tracking only $\sim$10 dark matter events are needed for discovery, versus of order $\sim$$(10^2-10^3)$ for experiments with no-HT, or axial, tracking \cite{MorganGreen20053D, MorganGreen20052D, MorganGreen2007, Copi2007, Billard2015}.  Studies of the ionization energy loss ($dE/dx$) of low energy recoils indeed predict a decrease with decreasing energy along the direction of the traveling recoil \cite{hitachiLET}.  Although a HT asymmetry of this nature has been measured down to about 40 keVr \cite{driftheadtail}, it appears to be small and after straggling and diffusion its detection is diminished to the point where the statistical advantage over the axial case is reduced.  Future experimental measurements are needed to determine the lowest energy at which this effect exists as well as its dependence on the recoiling atom and detection medium.

Based on the discussion thus far, it is clear that there is a large parameter space available for optimizing directional experiments to have the highest sensitivity, but with 
economical, robust and scalable designs.  As mentioned at the outset, in this work we defer the optimization discussion and focus instead on measuring the intrinsic properties of low energy electron and nuclear recoils.  For this goal, we made measurements with a small prototype TPC, which has the lowest diffusion that a realistic experiment could achieve, and chose a readout technology with an emphasis on high resolution and high signal-to-noise.  The detector, described in the next section, is based on Micro-patterned Gas Detector (MPGD) technology.  For a review of the development of MPGD technology, see Ref. \cite{sauliMPGD}.  Additionally, refer to Ref. \cite{RD51} for the webpage of the RD51 Collaboration whose goal is the advancement of MPGD technologies.  

In this paper we describe the detector performance, and using data from alpha, X-ray, gamma-ray, and neutron sources, we show how the salient features seen in the electron and nuclear recoil tracks can be used for background discrimination in dark matter searches.  In a second paper, we will describe the directional capabilities of the detector, which will be used to simulate its sensitivity to the directional signature from dark matter.

\section{Detector Setup}
\label{sec:Detector Setup}

The detector consisted of three standard copper GEMs (Gaseous Electron Multipliers \cite{sauli}; see Ref. \cite{buzulutskov} for a review) arranged in a cascade with 2 mm separation between them (Figure \ref{fig:detector}).  The GEMs were manufactured at CERN from $7$ $\times $ 7 cm$^2$ sheets of kapton (50 \si{\um} thick) clad on both sides by copper and mounted on G10 frames.  The surface of each sheet was chemically etched with biconical holes of diameter of 50/70 \si{\um} (inner/outer) configured in a hexagonal pattern with 140 \si{\um} pitch.  A cathode, placed 1 cm below the GEMs, was fabricated from a $7$ $\times $ 7 cm$^2$ copper mesh made from 140 \si{\um} wires with $\sim$320 \si{\um} pitch.  A 1 mm pitch anode wire grid plane made from 20 \si{\um} thick gold plated tungsten wires was located 3 mm above the top most GEM (GEM 3), forming the induction gap.  The detector was housed inside an aluminum vacuum vessel and calibrated using $^{55}$Fe (5.9 keV X-rays) and $^{210}$Po (5.3 MeV alphas) sources, both mounted inside the vacuum vessel.  A rotary feedthrough was used to individually turn both calibration sources on or off, as needed.  Before operating the detector the vacuum vessel was pumped down to $<$ 0.1 Torr and back-filled with 100 Torr of pure (99.999\%) $\text{CF}_{4}$ gas.  A BK-7 glass window was positioned above the readout grid to allow scintillation light from the final amplification stage (GEM 3) to be viewed by the CCD.  The BK-7 glass material was chosen due to its high transmittance of $\text{CF}_{4}$ scintillation, whose optical component is peaked around 620 nm \cite{morozov, kaboth}, and its lower cost relative to quartz.

A back-illuminated Finger Lakes Instrumentation (FLI) charge-coupled device (CCD) camera (MicroLine ML4710-1-MB) with a $1024 \times 1024$ pixel sensor array (CCD47-10-1-353) made by E2V was mounted on top of the vacuum vessel.  The 13 $\times$ 13 \si{\um}$^2$ square pixels occupy the 18.8 mm diagonal sensor, which has a peak quantum efficiency of 96$\%$ at 560 nm.  The camera could be read out at two speeds, 700 kHz and 2 MHz, with 16-bit digitization, and during data taking the sensor was cooled to the lowest stable operating temperature of \SI{-38}{\degreeCelsius} by a built-in Peltier cooler.  At this temperature, the read-out noise was measured at $\sim$10 e$^-$ rms and the dark current was $< 0.1$ e$^{-}$/pix/sec.  A fast 58 mm f/1.2 Nikon Noct-NIKKOR lens was mated to the CCD camera through a 20 mm extension tube for close-focusing imaging.  The CCD-lens system imaged a $2.8$ $\times$ 2.8 cm$^2$ region of the top most GEM surface.  The known pitch of the holes on this surface were used to calibrate the length-scale of the field of view.

\section{GEM Gain}
\label{GEM Gain}

A 5.9 keV $^{55}$Fe X-ray source was used to measure the effective gas gain, which includes the loss of electrons in the charge flow from the detection volume to the collection and readout surface.  With a W-value (the average energy per ionization) of $34.2$ eV in CF$_{4}$ \cite{christophorou}, the primary ionization from the 5.9 keV X-ray creates, on average, 172 electron-ion pairs per conversion event.  To measure the gas gain we used the standard procedure of using a pulser and capacitor to determine the preamplifier pulse height to charge calibration.  In our case, we used an ORTEC 448 research pulse generator and an ORTEC 142IH charge sensitive preamplifier, which comes with a built-in 1 pF capacitor for this purpose.  With the calibration results from this procedure the $^{55}$Fe energy spectrum obtained using a multi-channel analyzer (MCA) was used to determine the gas gain.  All gain measurements were made by reading out the signals from the last GEM electrode (GEM3 in Figure \ref{fig:detector}) with the preamplifier.  

We also attempted using the anode wire grid to read out the signal, as often found in the literature, but found that corona limited the maximum achievable gain.  The presence of the anode grid was not superfluous, however, as we found that the sparking probability tended to increase without it.  Reference \cite{bondar} has a discussion on the operation of multiple GEMs without the use of an anode board or wire plane for readout.  

\begin{figure}[]
 \centering 
	\includegraphics[width=0.45\textwidth]{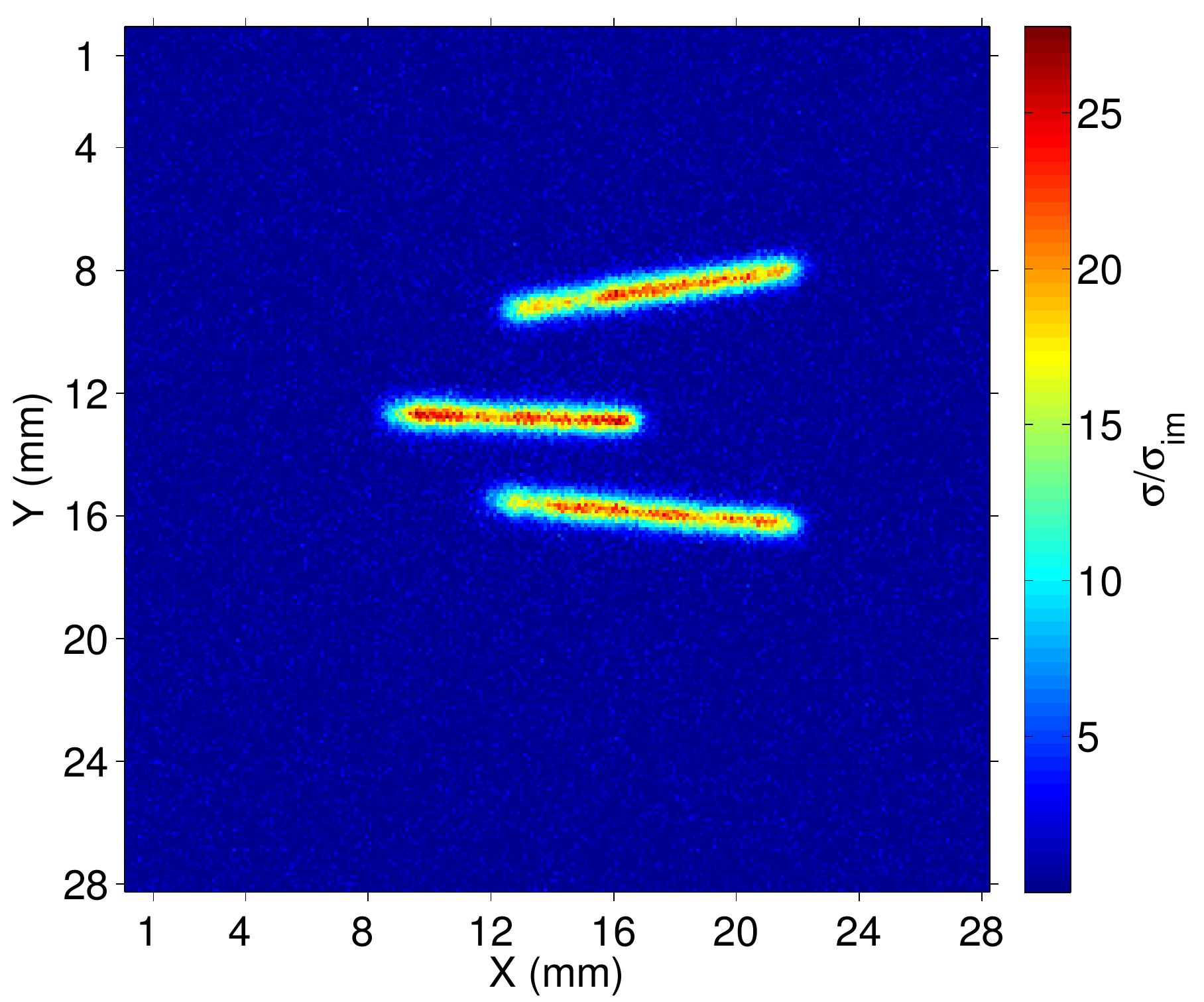}
  \caption[ ]{Image of alpha track segments in 100 Torr CF$_{4}$ with an effective gas gain of $1\times 10^{5}$ and $6\times$6 CCD binning. The image is contrast adjusted to show the signal to noise in units of $\sigma_{\text{im}}$, the rms of the image background.  The track segments ($\sim$1 cm) correspond to the parts of the tracks well before the Bragg peak and demonstrate the extremely high signal-to-noise level achieved with the brightest part of the tracks being $>20\sigma_{\text{im}}$ above the noise.}
 \label{fig:alphasegment}
\end{figure}

To optimize the GEM voltages, the detector was irradiated with alphas for about 1 hour ($\sim$500 alpha interactions) at each setting to test for stability.  If no sparks occurred during this time, then the detector was deemed to be stable.  The voltages were then changed and the procedure was repeated until the setting corresponding to the highest stable gain was found.  At a pressure of 100 Torr, with the biases of GEMs 1 and 2 $= 290$ V and GEM $3 = 460$ V, we obtained an effective gain $> 3\times 10^5$.  These settings corresponded to a drift field of 400 V/cm, transfer field of 1.45 kV/cm, and induction field of 315 V/cm.  The unbalanced powering scheme with different biases on the GEMs was used to circumvent the corona problem we experienced while operating in balanced power mode at low pressures (75 Torr and 100 Torr).  The disadvantage of such a power scheme with a large fraction of the gain coming from the last GEM is an increase in the sparking probability.  Indeed, sparking was observed with the $^{210}$Po alpha source (5.3 MeV) turned on when operating at a gain of $2.5\times10^4$ at 75 Torr and $3\times10^5$ at 100 Torr.  Although the sparks did not damage the GEMs or other detector components during the test runs (1 hour per voltage setting), they did saturate the CCD, producing an artifact known as residual images or ghost images \cite{janesick}.

Ghost images are often associated with front-illuminated CCDs, rather than back-illuminated CCDs, but were nevertheless observed in the CCD used in this work.  They appeared in frames taken after the initial saturation event and did not fade until many hours afterwards; in general, the relaxation time depends on the temperature of the CCD.  Even though they can be identified as spatially non-varying objects across successive frames, or can be dealt with by flooding the sensor with IR light, a technique known as RBI flushing, their presence is indicative of instability with the potential of damaging the detector.  Thus, we adjusted the GEM voltages to find the maximum gain attainable without sparking or corona at 100 Torr, to ensure stable long-term detector operation.  Stability was found with GEM $1 = 279$ V, GEM $2 = 334$ V, and GEM $3 = 380$ V, drift field of 400 V/cm, transfer fields of 1.40 kV/cm and 1.67 kV/cm between GEMs 1 and 2 and GEMs 2 and 3, respectively, and induction field of 260 V/cm between GEM 3 and the grid.  With these settings an effective gain of $\sim1\times10^5$ was achieved.  The excellent signal to noise achieved at this gain is illustrated by the alpha tracks in Figure \ref{fig:alphasegment}, which show a peak signal of $>20\sigma_{\text{im}}$ above the noise level.  Besides its use for gain measurements, the charge signal from the preamplifier was not used in the subsequent analysis of the data.  As the CCD camera was operated in non-trigger mode, it would be difficult to correlate the charge signal with a particular event in the CCD image.  Additionally, events with energies above three times the $^{55}$Fe energy would saturate the preamplifier.  However, using both the charge and light signals should further aid discrimination.

\section{Detector Calibrations}
\label{Detector Calibrations}

\subsection{CCD Calibration}
\label{CCD Calibration}

The standard approach to CCD calibration was adopted where each CCD image (or frame) was calibrated using a set of co-averaged flat-field and dark frames.  The flat-field frames were used to correct for vignetting and pixel to pixel variation in sensitivity, and the dark frames corrected for the variable accumulation rate of dark current across pixels.  The calibration was done by subtracting the co-averaged dark frame from each image frame and then dividing the resulting frame by the normalized, co-averaged flat-field.  Bias frames, which correct for the electronic bias that is seen as structure in the data frames, were not used since this information is also present in the dark frames.  The pedestal due to amplifier bias was removed using the overscan region present in each frame.  

With the CCD being read-noise limited, pixels were binned $6 \times 6$ prior to digitization and read out at the slowest allowable speed of the CCD electronics, 700 kHz, to improve signal to noise.  This binning factor combined with the imaging area translated to each binned pixel imaging a $\sim$165 $\times$ 165 \si{\um}$^2$ area of the  
GEM, which is well-matched to the 140 \si{\um} GEM pitch.  The measured read noise was 10 e$^{-}$ rms/pix, and at the ${-38}^{\circ}$C operating temperature of the CCD, the dark current was $\sim$0.03 $e^{-}$/pix/sec for $1\times1$ binning.  As the latter scales with pixel area, the $6 \times 6$ binning and 5 second exposures used in all of our data runs contributed a dark current of $\sim$6 e$^{-}$  to the total system noise.

The calibration frames were averaged using an algorithm that rejects pixels hit by cosmic rays and radioactivity by comparing the value of the same pixel across each frame, and rejecting those above three sigma of the initial average of these pixels.  The average value of the pixels was then recalculated excluding the rejected pixels and the same procedure repeated until convergence in the average was reached.  This procedure was applied to all pixels to create master flat and dark calibration frames.

\begin{figure}[]
 \centering 
    \includegraphics[width=0.45\textwidth]{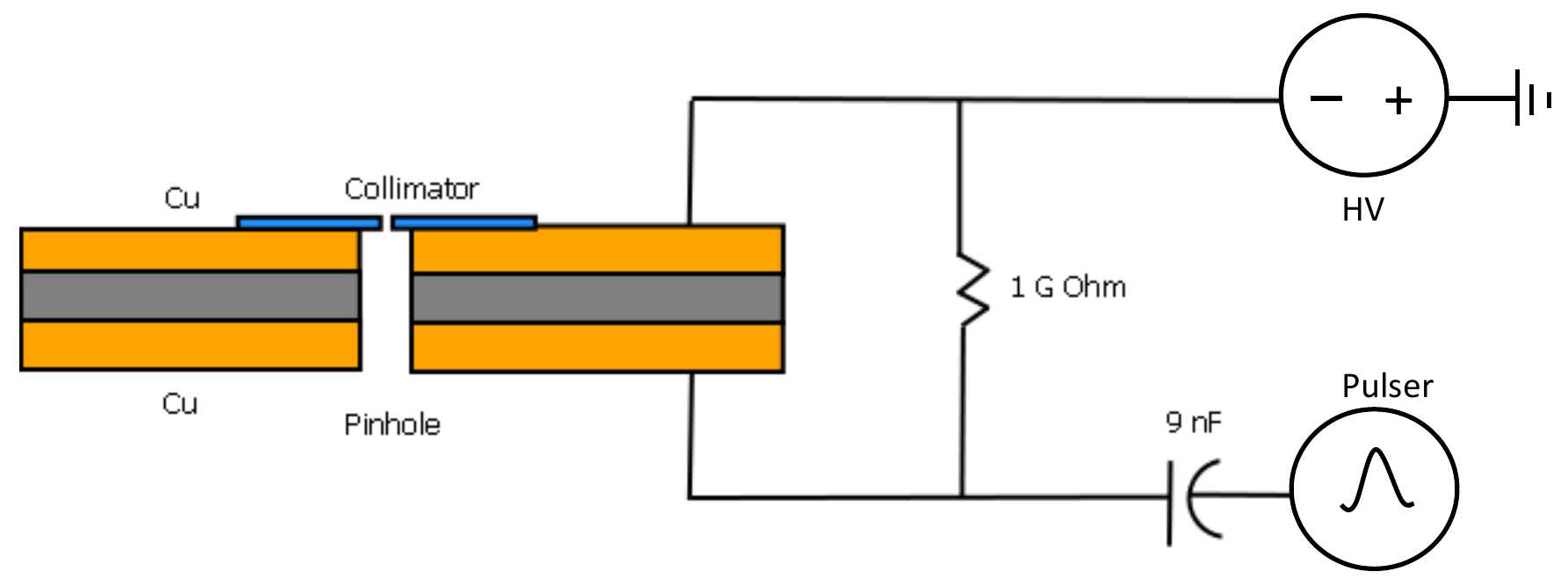}
  \caption[ ]{A schematic of the pinhole cathode and high voltage pulser setup used to measure transverse diffusion.  Both surfaces of the cathode are held at a fixed voltage with a high voltage power supply.  The capacitor isolates the pulser from the high voltage but allows an impulse to be transferred through.  The resistor (1 \si{\giga\ohm}) and high voltage impulse (12 kV) causes a short duration voltage change only on the bottom surface of the device (the top surface is held a fixed voltage by the power supply), initiating a spark in the pinhole and generating ionization.  The 10 \si{\um} collimator reduces the transverse extent (perpendicular to electric field) of the ionization entering the drift volume (region on top of the collimator) to point-like dimensions. }
  \label{fig:cathodepinhole}
\end{figure}

\subsection{Transverse Diffusion}
\label{Transverse Diffusion and Spread in Track}

The detector track reconstruction and angular resolution is intrinsically tied to the diffusive properties of electrons in the target gas.  With our CCD readout, only the transverse, or lateral, diffusion could be measured as the longitudinal component required timing resolution from the GEM charge readout that is well beyond the capability of the ORTEC preamplifier used here.  To measure the lateral diffusion, a point source was generated at a specially constructed cathode made from an insulating sheet sandwiched between two strips of copper tape.  The cathode had a small hole punctured at its center with a 10 \si{\um} collimator placed over the hole on the side facing the drift volume.  The electrodes of the cathode were connected to a power supply and a 12 kV high voltage pulse generator; Figure \ref{fig:cathodepinhole} shows a schematic of the cathode and electrical connections.  The HV pulser was used to generate a spark inside the hole, which produced ionization that appeared as a point source `track' in the drift volume after passing through the collimator.\footnote{As a check on this technique for generating point tracks, we also used alpha track segments (e.g., Figure \ref{fig:alphasegment}) and found that measurements of their widths gave results in good agreement.}

We imaged a collection of tracks and measured the spread in their light profiles.  Each of the light profiles was fitted using a Gaussian curve, and an average $\sigma_{\text{tot}} = 345$ $\pm$ 5 \si{\um} (stat) was obtained for the sample of tracks.  The main contribution to the spread is expected from diffusion of the electron cloud as it travels the full distance from the cathode to the final GEM stage.  This was confirmed using data from Refs. \cite{Christophorou1, Christophorou2} and results from the MAGBOLTZ program \cite{biagi}, which predict $\sigma_{\text{diff}}= 326$ \si{\um} for the diffusion in our detector.  The MAGBOLTZ result indicates that contributions to $\sigma_{\text{diff}}$ are divided about equally, when added in quadrature, between the drift gap and the two transfer gaps between the GEM stages.  A secondary source to $\sigma_{\text{tot}}$ is expected from the GEM pitch and the 3-GEM cascade.  Attributing this wholly to the difference between $\sigma_{\text{tot}}$ and $\sigma_{\text{diff}}$ yields 65 $\pm$ 9 \si{\um} per GEM when added in quadrature.  This value is consistent with our expectation that the track spread due to the GEM pitch is roughly half of its 140 \si{\um} pitch.  We expect other contributions, such as smearing due to imperfect optics, to be minor compared to these.  Although quite low in our prototype detector, diffusion could be further reduced if the transfer gap contribution could be eliminated, or if a negative ion drift gas such as the CS$_2$ could be employed.  With diffusion scaling as $\sqrt{L}$ with drift distance, these considerations become critical for scale up to large detectors; this is discussed further in Section \ref{sec:detector improvements}.

\subsection{Energy Calibration}
\label{Energy Calibration}

\begin{figure}[]
 \centering 
    \includegraphics[width=0.45\textwidth]{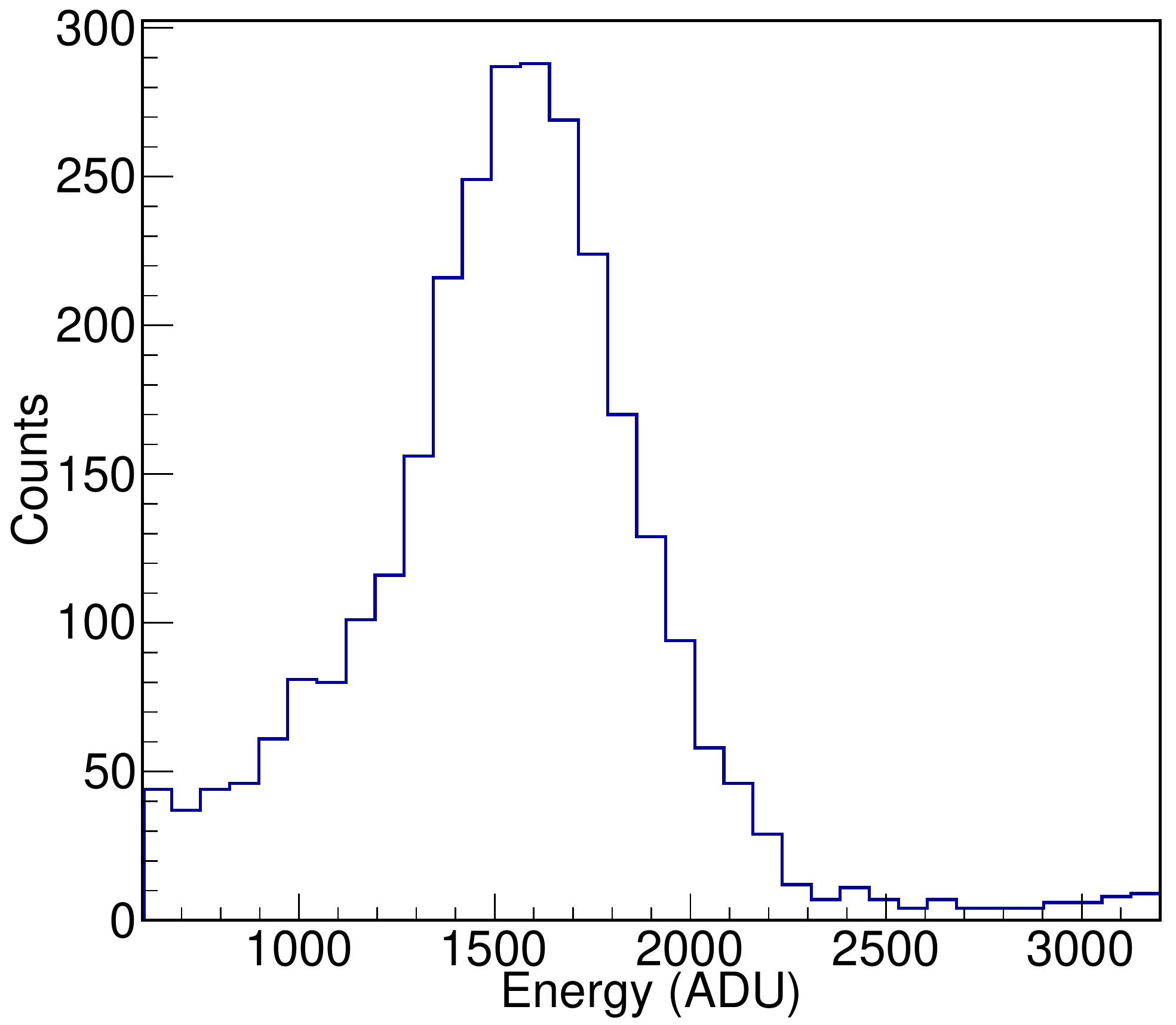}
  \caption[ ]{An $^{55}$Fe energy spectrum obtained with the CCD camera and used to calibrate the energy of recoils for one of the sub-runs.  The energy is shown in ADUs (analog-to-digital-units).  The conversion factor is found by taking the ratio of the fitted peak of the spectrum to the average energy of an $^{55}$Fe recoil (5.9 keV).}
  \label{fig:fe55spectrum}
\end{figure}

Energy calibrations were done using $^{55}$Fe X-ray and $^{210}$Po alpha sources.  The alpha track calibration was made by first using SRIM \cite{ziegler} to calculate the Bragg curve of a 5.3 MeV alpha in 100 Torr CF$_{4}$.  We measured the location of the alpha source relative to the drift volume and determined the part of the track that would be imaged by the CCD camera.  Figure \ref{fig:alphasegment} shows segments of alpha tracks imaged by the CCD camera at the maximum stable gain.  By comparing the total integrated light output in the image of the alpha track with the energy calculated from the SRIM generated Bragg curve, we obtained the light to energy conversion factor, ADU/keV$_{\alpha}$.  Since $>$ 99$\%$ of the energy lost by an alpha particle before its Bragg peak is through ionization, we can treat this as keV electron-equivalent energy (keVee). 

\begin{table}[]
\caption{Detector Parameters}
\label{tab:params}
\centering
\begin{tabular}{ |p{4.5cm}|p{2.8cm}|  }
 \hline
 \multicolumn{2}{|c|}{\textbf{Detector Parameters}} \\
 \hline
\multicolumn{2}{|l|}{CCD and Imaging Parameters} \\
\hline
 Peak QE  & 	96\% (560 nm)     		\\
 Pixel Size	($1 \times 1$ binning)			&	$13\times13$ \si{\um}$^{2}$	 				\\
 Pixel Binning   				&	$6\times6$					\\
 Binned Pixel Imaging Scale  &  165 \si{\um}/pix \\
 Imaging Area	&   $2.8\times2.8$ cm$^2$    		\\
 Read Noise	@ 700 kHz		&   10 e$^-$ rms		\\
 Operating Temperature  & $-38^{\circ}$ C \\
 Dark Current				& 	0.03 e$^{-}$/s/pix   				\\
 Exposure Time					& 	5 sec  					\\
\hline
\multicolumn{2}{|l|}{Vessel Parameters} \\
\hline
 Detection Volume			& 	$2.8 \times 2.8 \times 1.0$ cm$^3$ 					\\
 CF$_4$ Pressure			& 	100 Torr  				\\
 Effective Gas Gain				& 	$10^5$ 					\\
 Effective Transverse Diffusion				& 345 $\pm$ 5 \si{\um} 					\\
 \hline
\end{tabular}
\end{table}

For an independent calibration method we imaged the electronic recoils from $^{55}$Fe X-ray interactions and obtained an energy spectrum of the scintillation signal; see  Figure \ref{fig:fe55spectrum}.  At our maximum stable effective gain, $^{55}$Fe tracks were visible at $6 \times 6$ pixel binning with a FWHM energy resolution of 38$\%$.  To our knowledge, this represents the first optically obtained spectrum of $^{55}$Fe in a TPC detector (details provided elsewhere).  The two calibration methods give results that are within 20\% of each other.  The small difference could be due to a systematic in determining the alpha segment imaged by the CCD camera.  In the analysis of the data to follow, we will use the energy conversion factor derived from the $^{55}$Fe energy calibration.  Lastly, a summary of important detector parameters discussed thus far is included in Table \ref{tab:params}.

\section{$^{60}$Co and $^{252}$Cf Data Runs}
\label{sec:Co-60 and Cf-252 Runs}

To study our detector's response to nuclear recoils, and its ability to distinguish these from gamma backgrounds, we used a $^{252}$Cf neutron source and a $^{60}$Co gamma source.  For the $^{60}$Co run, the source was placed outside the vacuum vessel but inside a lead housing to protect the CCD sensor from direct gamma-ray interactions; there was no lead between the source and the outer vessel wall.  The neutron run was conducted in a similar manner but with the addition of lead bricks between the source and detector to attenuate the large number of gammas from $^{252}$Cf.  In total, about 96 and 36 hours of neutron data and gamma data were collected, respectively.  To evaluate the detector's directional sensitivity half of the neutron data was collected with the neutrons directed in the $-x$ direction, an axis lying in the imaging plane, and the other half with the neutrons directed in the $+x$ direction.  

For each data taking sequence, or sub-run, the vessel was pumped out and back-filled with fresh CF$_{4}$ gas to a pressure of 100 $\pm$ 0.05 Torr and sealed.  This was followed by powering up the GEMs to the voltage settings corresponding to maximum stable gain (see Section \ref{GEM Gain}).  The CCD pixels were binned $6\times6$ on-chip and the chip cooled to the lowest stable operating temperature of \SI{-38}{\degreeCelsius}, which was monitored by an internal sensor in the camera.  An energy calibration was done with an $^{55}$Fe source at the start and end of each sub-run sequence.  The high drift speed of electrons in CF$_{4}$ made it impossible to trigger the CCD (open and shut the shutter) using the charge signal from the first GEM stage.  Therefore, we operated in non-trigger mode with the CCD camera successively taking 5 second exposures over a duration of about 12 hours for each sub-run.  This corresponded to approximately 9 hours of live time after accounting for the CCD readout time.  The detector was refilled after each sub-run to avoid substantial gain degradation due to changes in gas purity from out-gassing during the data taking sequence.  As this and other effects including temperature and composition changes from charge avalanching caused the gain to drift over time, we used the average of the values measured at the end and start of each data sequence for the energy calibration constant.  In total, eight sub-runs were conducted for the neutron data and three for the gamma data.  

We analyzed the data using an image analysis algorithm developed with MATLAB and its image processing toolbox.  First, images were calibrated and binned $4\times4$ in software, and pixels above a set threshold of 3.2$\sigma_{\text{im}}$ were identified as objects.  All objects found crossing the image boundaries were rejected.  
The binned image was then up-sampled back to its original size, resulting in an index image for the pixel locations of all identified objects.  In the remainder of the analysis, all object properties were determined from the original, non-software binned image.  To exclude hot pixels and CCD events (objects resulting from direct interactions of cosmic rays, radioactivity, neutrons, or gamma rays with the CCD sensor), we required objects to contain at least four contiguous pixels.  Separated pixels belonging to a local grouping of pixels above threshold were connected back to the primary grouping by morphologically closing the object using a disk-shaped structuring element with a radius of two pixels.  In essence, the closing operation, which is a dilation follow by an erosion, connected all pixels above threshold that lay within the radius of the structure element.  Each identified object was fitted with a position and an intensity weighted ellipse, which, along with the pixel grouping in the unfitted object, were used to determine some of its important properties such energy, track length, width, skewness, and energy loss profile.

\section{Results}
\label{Results}

\subsection{Background Discrimination}
\label{sec:Background Discrimination}

\begin{figure*}[]
	\centering
	\subfloat[$^{60}$Co data pre-cuts]{\includegraphics[width=0.45\textwidth]{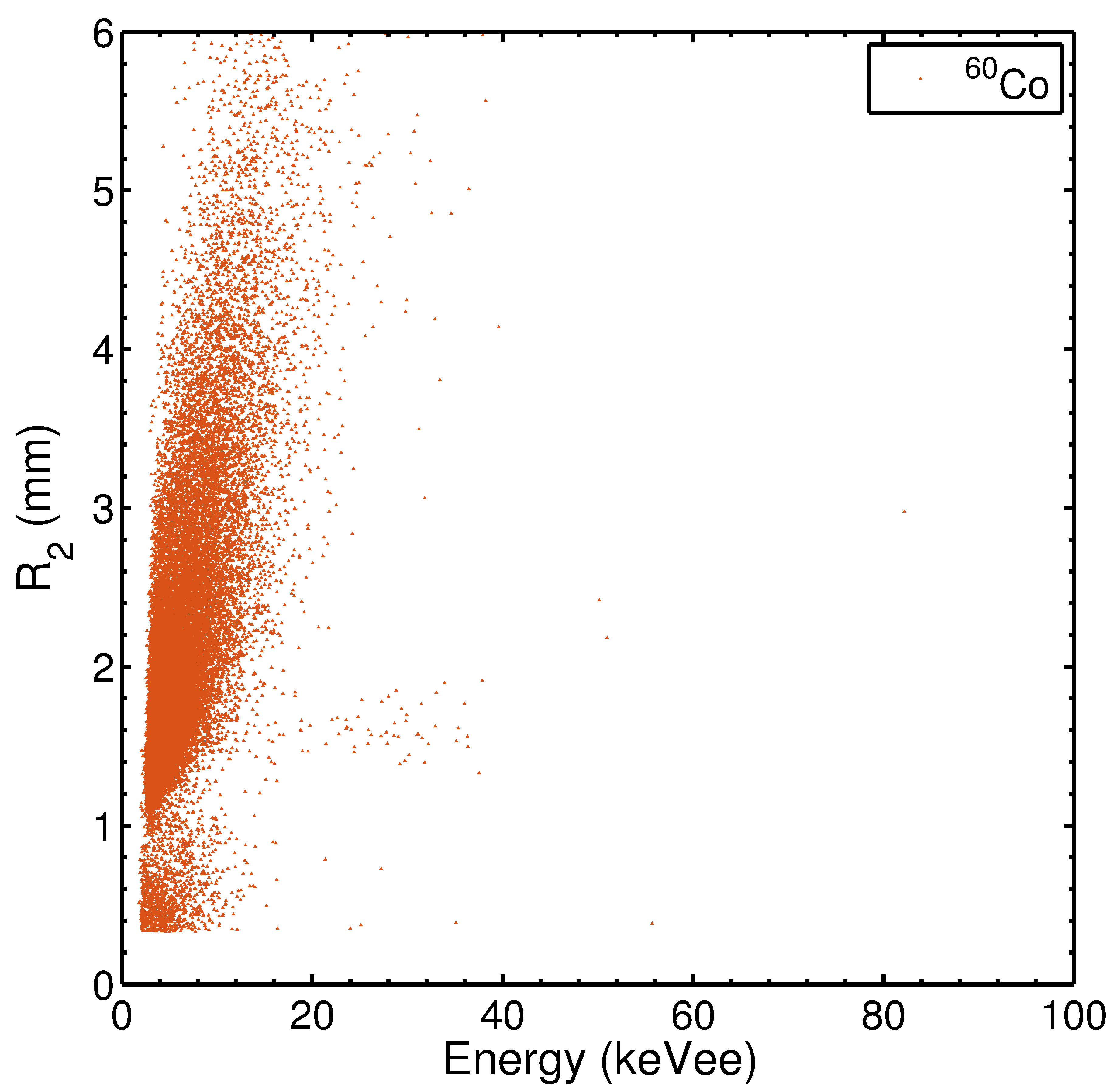}
	\label{fig:gamma1}}
	\qquad
	\subfloat[$^{60}$Co data post CCD cuts]{\includegraphics[width=0.45\textwidth]{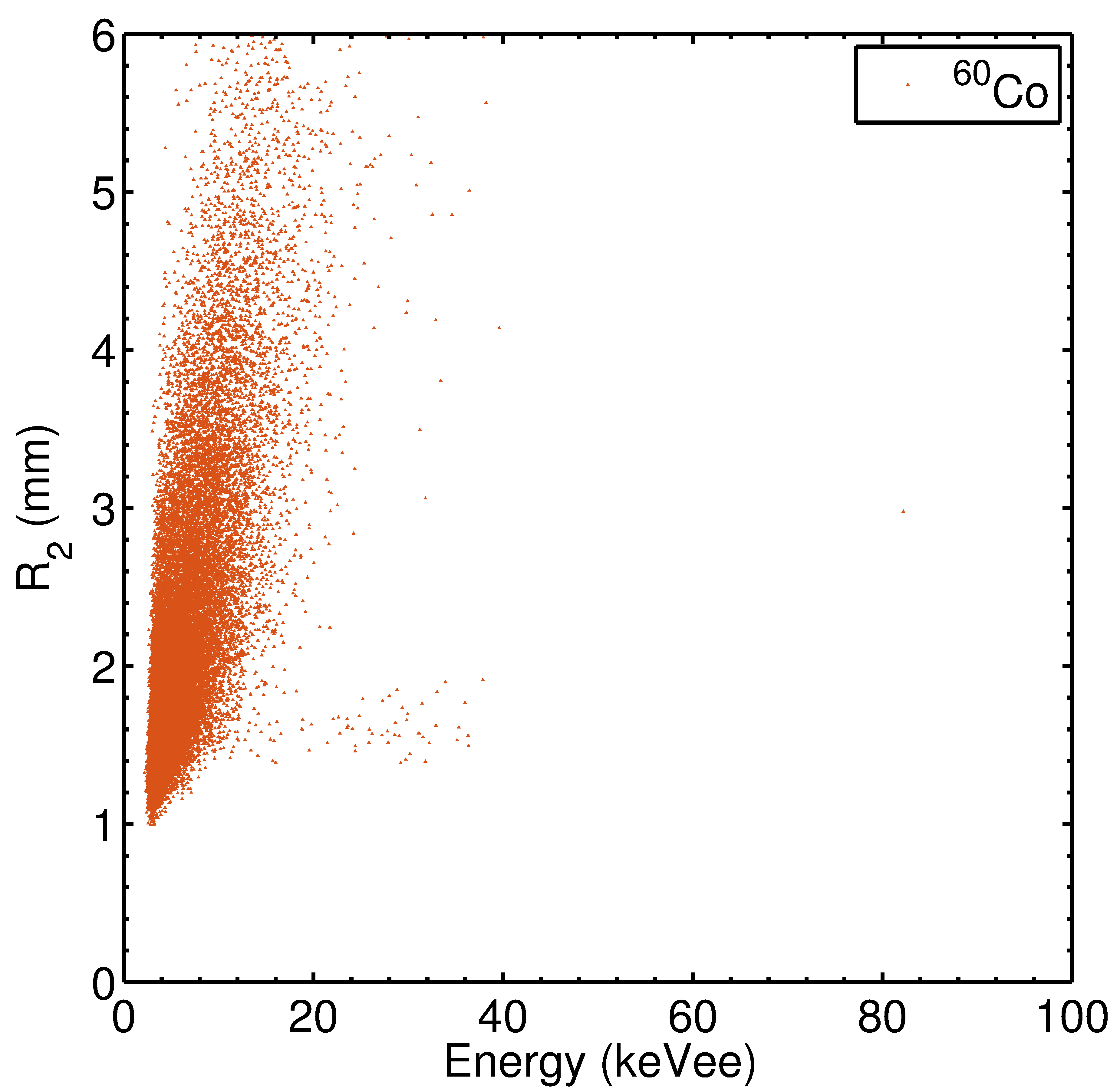}
	\label{fig:gamma2}}
	\caption[]{(a) The projected range ($R_{2}$) vs. energy plot of data from the $^{60}$Co gamma run.  The events with short range and low energy in the lower left corner of the plot are called CCD events, which result from the direct interaction with the imaging sensor. (b) The same data after analysis cuts are made to remove the CCD events.  The events in the short horizontal band extending to 40 keVee lie in the nuclear recoil band (see Figure \ref{fig:neutron12} and text).  These events are likely due to radon progeny recoils occurring at the cathode or GEM surfaces. }
	\label{fig:gamma12}
\end{figure*}

\begin{figure*}[]
	\centering
	\subfloat[$^{252}$Cf data post CCD cuts]{\includegraphics[width=0.45\textwidth]{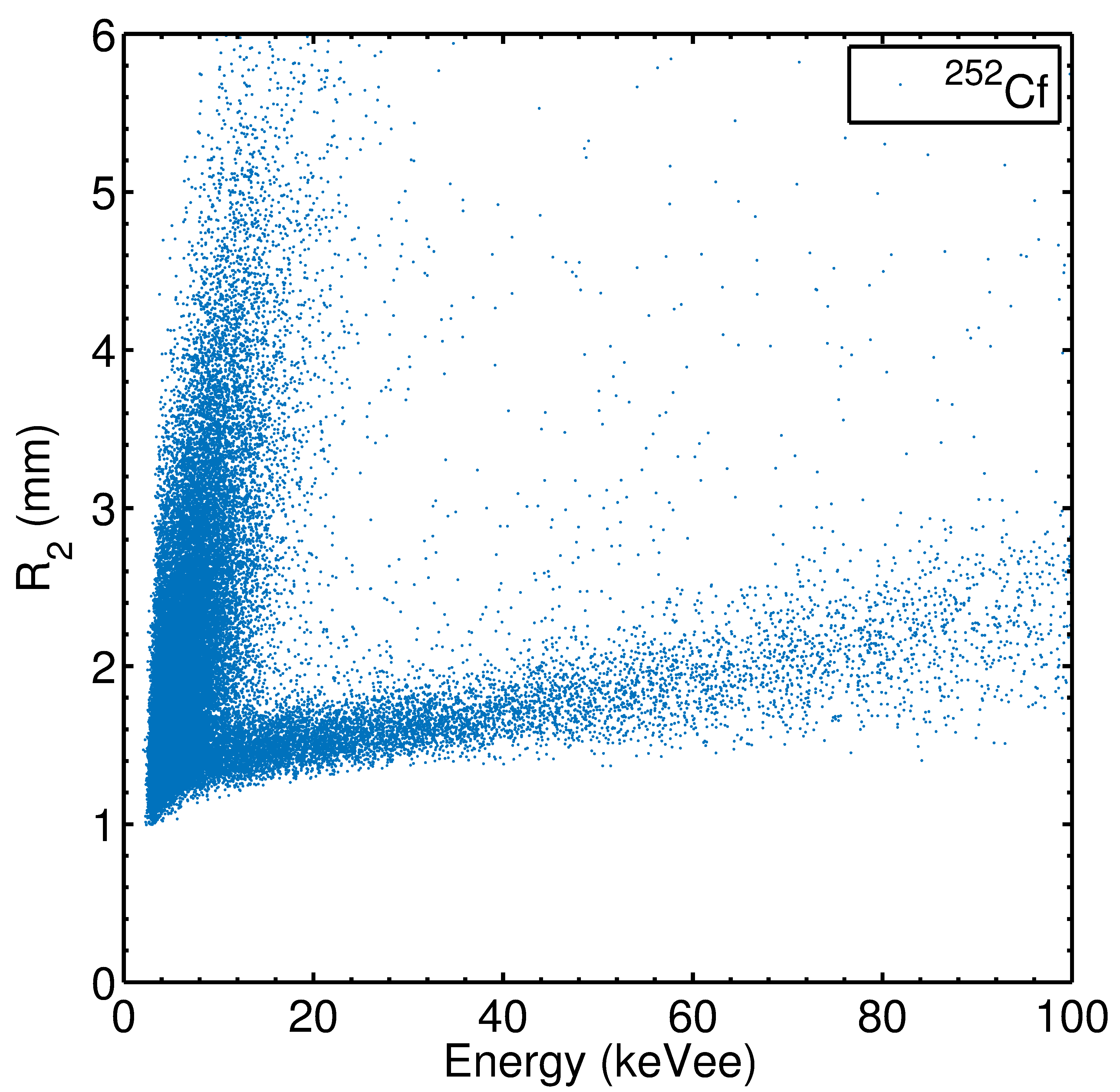}
	\label{fig:neutron1}}
	\qquad
	\subfloat[$^{252}$Cf data post selection cuts]{\includegraphics[width=0.45\textwidth]{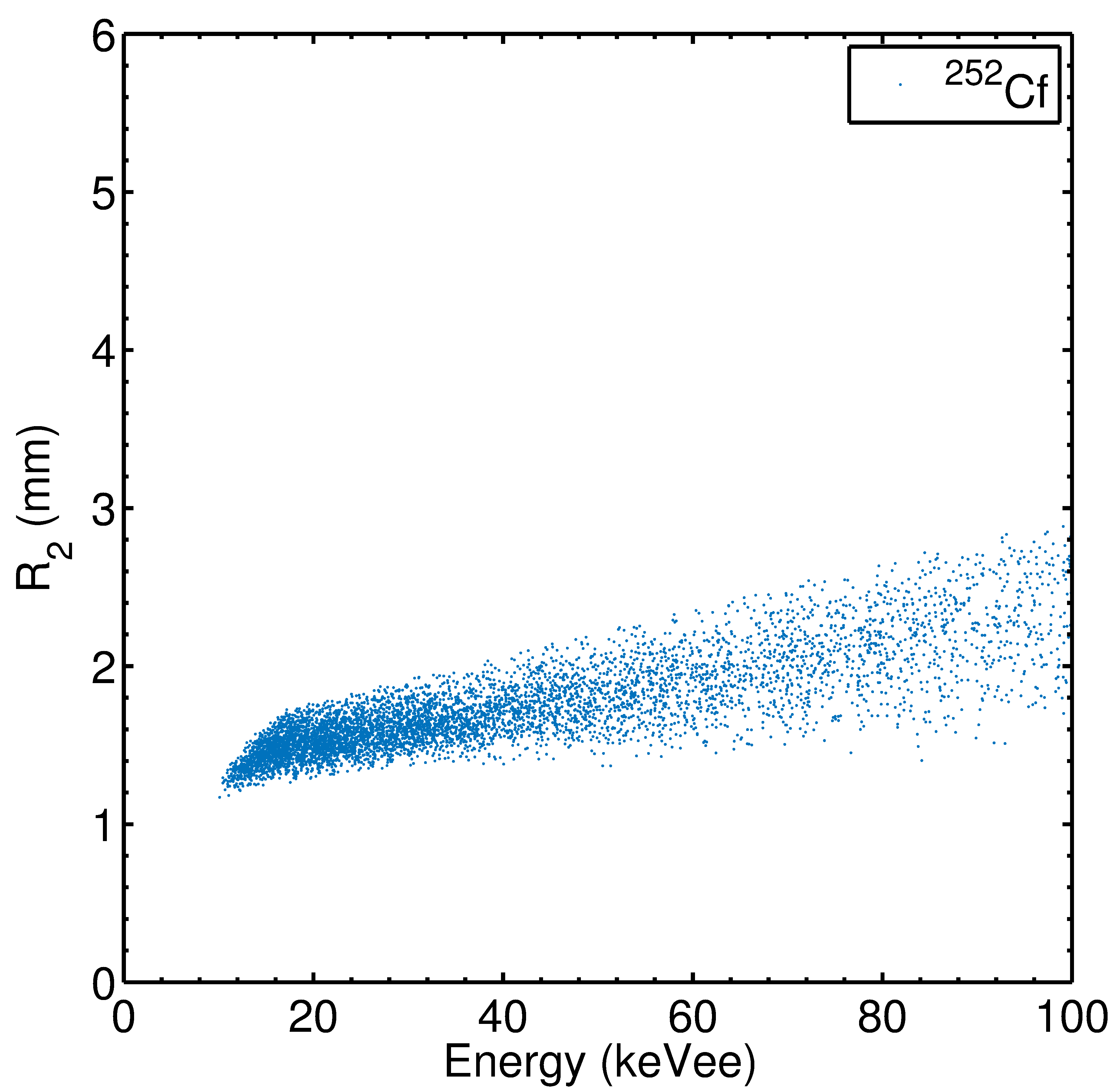}
	\label{fig:neutron2}}
	\caption[]{(a) The projected range ($R_{2}$) vs. energy plot of data from the $^{252}$Cf run after applying the CCD event cuts.  The events not part of the two bands are most likely segments of proton recoils created by neutron interactions with hydrogen-rich materials in the detector. (b) The same data with nuclear recoil selections cut applied.  The lowest energy recoils post-cuts extend to $\sim$10 keVee (23 keVr). }
	\label{fig:neutron12}
\end{figure*}

The background rate in a detector can vary widely depending on its size and the materials used in its construction, with even the most stringent requirements on radio-purity not eliminating all sources of backgrounds.  Consequently, fiducialization and discrimination are of critical importance to dark matter and other rare event searches that require large detection volumes.  One important source of backgrounds are gamma-rays and X-rays that can interact inside the detector to produce electronic recoils.  For tracking detectors, the stopping power, $dE/dx$, provides a powerful tool for discriminating between electronic and nuclear recoils.  Electronic recoils have a much lower average $dE/dx$ and, hence, much longer ranges as compared to nuclear recoils of the same energy, a fact that is evident in the range versus energy plots shown in Figures \ref{fig:gamma12} and \ref{fig:neutron12}.  Given the inability of our detector to measure the Z-component of the track, the range in these figures is 2D.  Of course, to maximize separation between the nuclear and electronic recoil bands, a detector with full 3D tracking capability is desirable (see Section \ref{sec:Background Discrimination 1D 2D 3D}).

\subsection{Gamma and Neutron Data}
\label{sec:Gamma and Neutron Data}

Using the reconstructed tracks passing the track identification algorithm from the $^{60}$Co gamma run, the 2D range as a function of energy is shown in Figures \ref{fig:gamma1} and \ref{fig:gamma2}.  The hard vertical edge at 2 keVee is the result of a software threshold set on the energy of detected objects to reduce the number of false event detections during the initial track finding stage of the analysis.  The sub-mm events in the lower left region of Figure \ref{fig:gamma1} are the CCD events described in Section \ref{sec:Co-60 and Cf-252 Runs}, which are due to direct interactions of ionizing radiation with the CCD sensor.  As these CCD events suffer no diffusion, they tend to have extremely high standard deviations of their pixel values as well as very high average intensities (total intensity/number of pixels).  Therefore, cuts made on these two parameters, in addition to track size, were used to efficiently remove this class of events.  The events in the small branch protruding from the primary vertical band at around 1.5 mm are mostly due to detector intrinsic backgrounds from decays of radon daughters.  These are sometimes referred to as radon progeny recoils, or RPRs, and occur at the detector surfaces \cite{miller1, miller2}.  Events in the primary vertical band have low average $dE/dx$ and correspond to electron recoil events.  These are primarily due to Compton scatterings of the 1.17 and 1.33 MeV gamma rays emitted in the beta decay of $^{60}$Co, with a small fraction from ambient and intrinsic electromagnetic backgrounds in the detector.  Altogether, there were $27,644(25,761)$ events from the gamma run before(after) applying the selection cuts to remove CCD events. 

The same plots for reconstructed tracks from the $^{252}$Cf neutron run are shown in Figures \ref{fig:neutron1} and \ref{fig:neutron2}.  Two distinct bands are present.  The vertical band is the same electronic recoil band observed in the gamma run, while the second, near-horizontal band contains the high $dE/dx$ nuclear recoils (both carbon and fluorine with a ratio implied by a GEANT simulation of 1:6).  The events forming the ``haze"  between these two primary bands have $dE/dx$ values inconsistent with being due to Compton scatters, and they were absent in the $^{60}$Co data.  Their  $dE/dx$ is also inconsistent with those of carbon or fluorine recoils as their lengths far exceed the maximum of these recoiling ions at a given energy.  Since these events were only seen in the neutron runs, we believe that they are segments of proton recoils from neutron interactions with hydrogen rich material in the detector such as the GEM kapton substrate.

\begin{figure*}[]
	\centering
	\subfloat[Discrimination parameter histogram]{\includegraphics[width=0.47\textwidth]{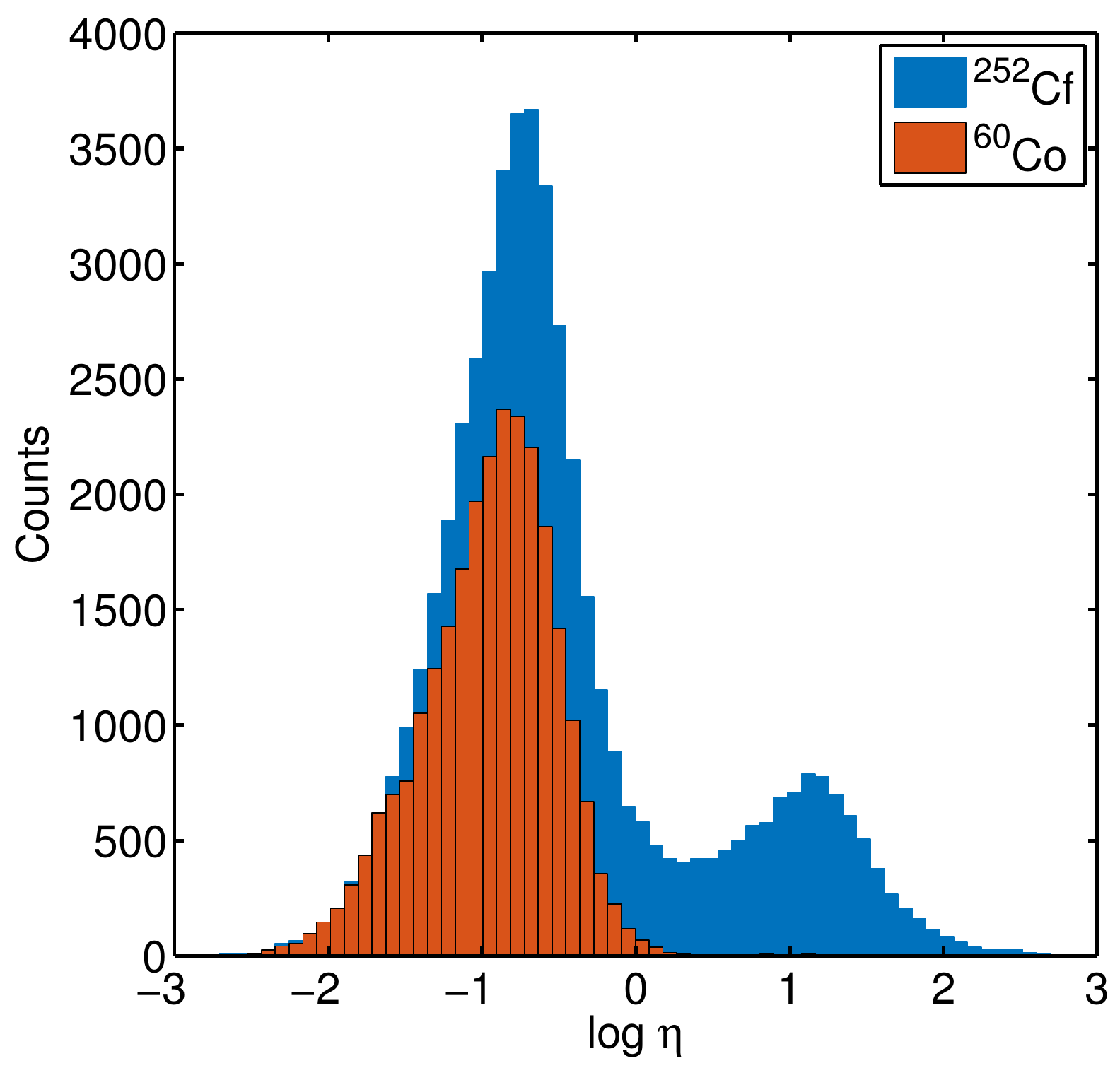}
	\label{fig:logetaneutrongamma}}
	\qquad
	\subfloat[$^{60}$Co data RPR separation]{\includegraphics[width=0.4475\textwidth]{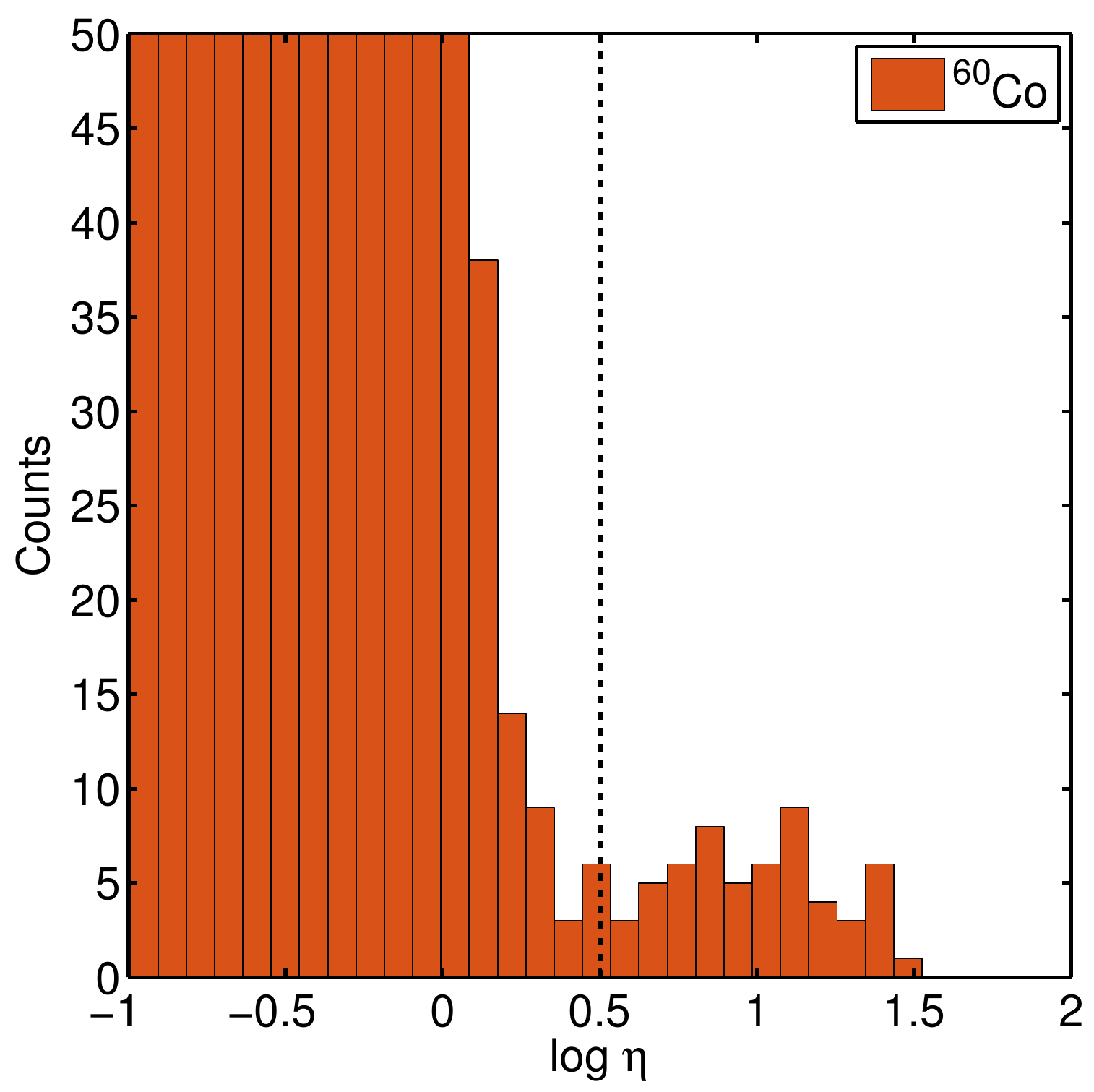}
	\label{fig:logetagamma}}
	\caption[]{(a) A histogram of the discrimination parameter, log $\eta$, defined as the ratio of the Bragg curve peak to track length, for the $^{60}$Co (red) and $^{252}$Cf (blue) runs.  There are two distributions in the $^{252}$Cf data representing the nuclear and electronic recoils while only one prominent one in the $^{60}$Co data which contains the electronic recoils.  (b) A histogram of the log $\eta$ parameter for the $^{60}$Co data zoomed in to see the small distribution of RPR events which overlaps the nuclear recoil peak in the $^{252}$Cf data.  The vertical dotted line at 0.50 is the value of the cut set on this parameter used for discrimination.}
	\label{fig:discrimparam}
\end{figure*}

From Figure \ref{fig:neutron1}, it is evident that even before any selection criteria are applied, there is good separation of nuclear recoils from electronic recoils down to low energies.  Nevertheless, we used the $^{60}$Co data to develop an algorithm that maximizes the rejection of electronic recoils while retaining a high detection efficiency for nuclear recoils.  One parameter that gave good separation between the two recoil classes is the ratio of the projected Bragg curve peak (peak of the light distribution in keVee) to the track length (major axis of the fitted ellipse in mm).  A histogram of the natural logarithm of this parameter, defined as $\eta$, is shown in Figure \ref{fig:discrimparam} for both the  $^{60}$Co and $^{252}$Cf data.  The distribution of the former has mainly one population whereas the latter, which contains both electronic and nuclear recoils, has two.  The one with the larger $\log \eta$ corresponds to nuclear recoils, as also confirmed by noting that they lie in the nuclear recoil band in the $R_{2}$ versus energy plots (Figures \ref{fig:neutron1}).  Based on these distributions, we defined the gamma cut of $\log \eta  <  0.50$ to reject electronic recoils.

Applying this cut to the gamma run eliminated all but 65 events, of which 56 have energies below 38 keVee and 9 have energies between 110 and 440 keVee.  We show below that these events are nuclear recoils and, in particular, the 56 lower energy events are consistent with RPRs from alpha decays occurring at detector surfaces \cite{miller1, miller2}, such as the cathode or GEM, while the remaining 9 of higher energies are probably segments of alpha tracks associated with those decays.  The ratio is not 1:1 because the alphas have much higher energies, and hence, greater probability to cross the image frame edge and be rejected by the analysis. 

\begin{figure*}[]
	\centering
	\subfloat[$^{60}$Co and $^{252}$Cf data post selection cuts]{\includegraphics[width=0.45\textwidth]{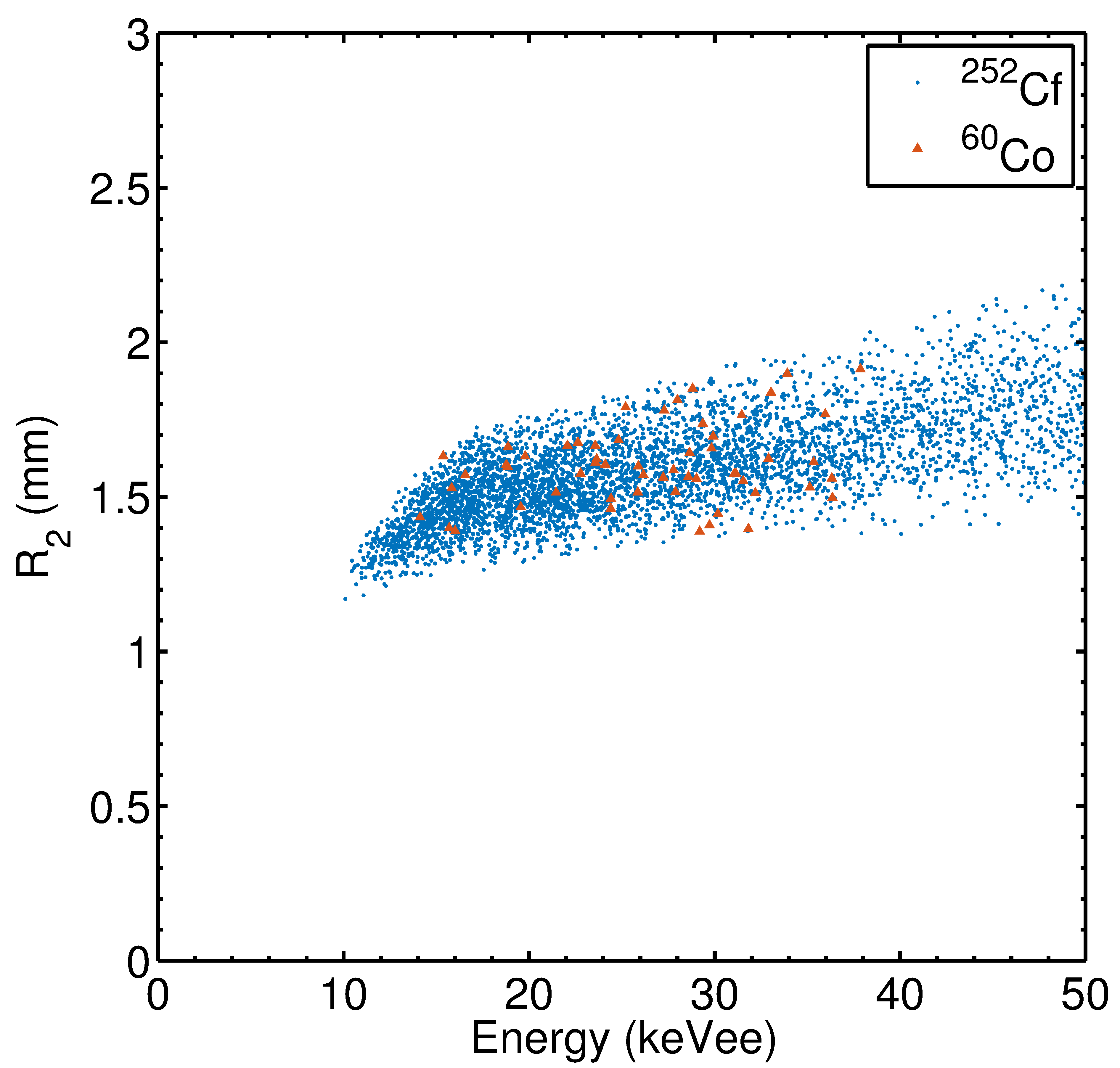}
	\label{fig:RPR_R2}}
	\qquad
	\subfloat[Recoil Energy Spectrum]{\includegraphics[width=0.45\textwidth]{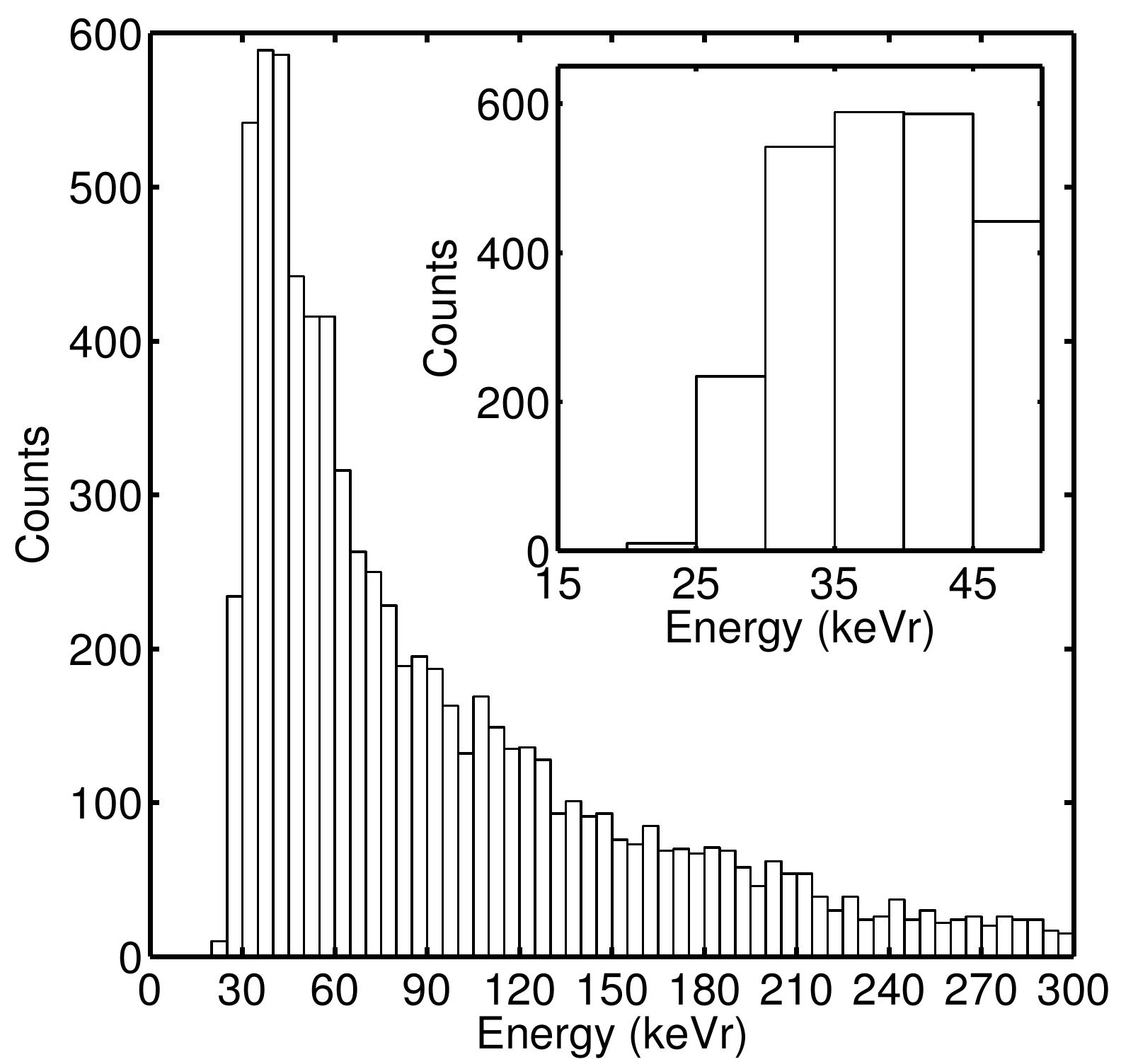}
	\label{fig:recoilspectrum}}
	\caption[]{(a) The R$_2$ vs energy plot for events passing the analysis cuts for nuclear recoils from both the $^{60}$Co and $^{252}$Cf datasets.  The events remaining from the $^{60}$Co data lie in the nuclear recoil band and have an energy distribution consistent with RPRs, as discussed in the text.  (b) The measured energy spectrum of carbon and fluorine recoils from $^{252}$Cf neutrons after analysis cuts were applied to remove the electronic recoils.  The spectrum, which is peaked around 35 keVr, was derived assuming the quenching factors from \cite{hitachi} and that all recoils were fluorine. }
	\label{fig:spectrum-keV}
\end{figure*}

There are several pieces of evidence that support the RPR interpretation.  First, a zoom-in of the $\log \eta$ histogram of the $^{60}$Co data, shown in Figure \ref{fig:logetagamma}, clearly shows the 65 events in question forming a distinct population, which overlaps that of the nuclear recoil events from the $^{252}$Cf data in Figure \ref{fig:logetaneutrongamma}.  That these events are nuclear recoils is further corroborated by the fact that they lie in the nuclear recoil band in the $R_{2}$ versus energy plane, as shown in Figure \ref{fig:RPR_R2}.  Finally, the evidence that these events are mostly, if not all, due to RPRs comes from their energy distribution.
For the 56 lower energy events this distribution has a mean of 26 keVee with a cut-off at 38 keVee, consistent with the ionization energy distribution from the short-lived lead isotopes in the radon chain \cite{MIMAC-RPRs}.

After excluding the 65 RPR and associated alpha events, the $^{60}$Co dataset contains 25,696 events, which we identify as electronic recoils.  Applying the gamma cut described above removes all of these, resulting in the detector's gamma rejection at $\le 3.9\times10^{-5}$.  This rejection level is achieved at a pressure of 100 Torr with two dimensional (2D) track reconstruction, and may be improved further with full 3D reconstruction (Section \ref {sec:Background Discrimination 1D 2D 3D}) and/or by operating at lower pressure, where tracks are longer and better resolved.  Additionally, more sophisticated analysis algorithms should give better results.

\begin{figure*}[]
	\centering
	\subfloat[ ][9 keVee  electron recoil]{
	\includegraphics[width=0.40\textwidth]{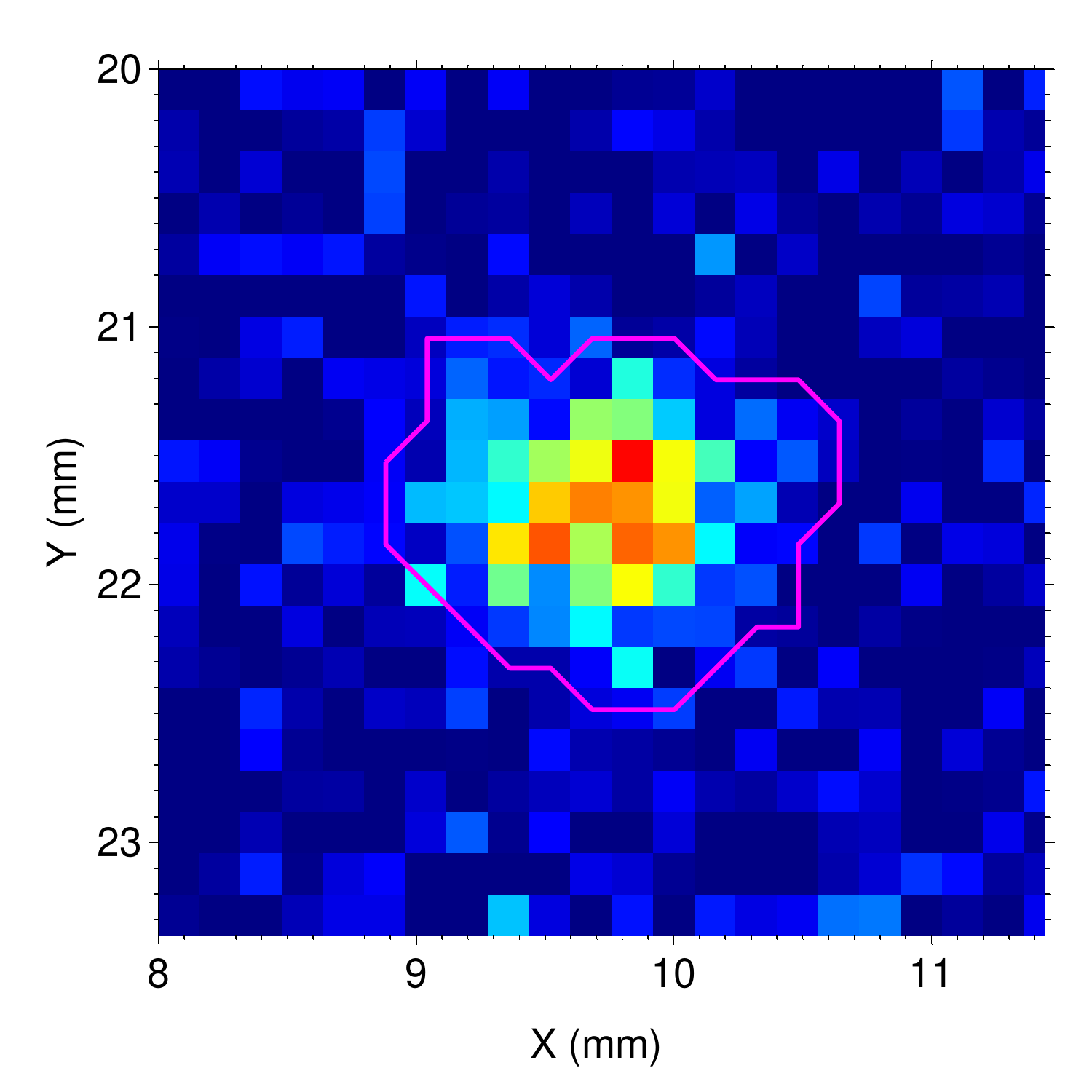}  
	\label{fig:electronic1}}
	\qquad
	\subfloat[ ][13 keVee  electron recoil]{
	\includegraphics[width=0.40\textwidth]{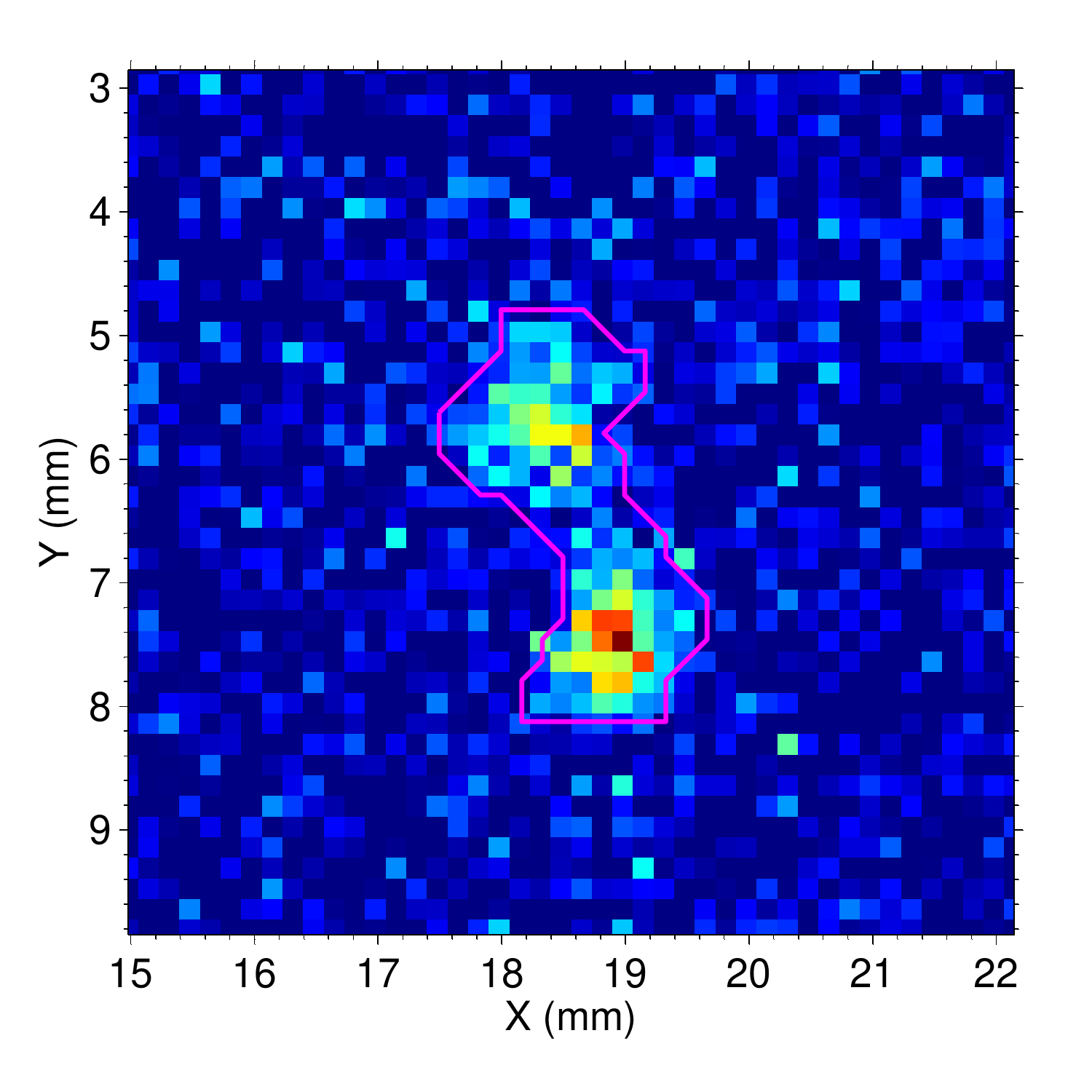}
	\label{fig:electronic2}}	
	\qquad
	\subfloat[ ][21 keVee electron recoil]{
	\includegraphics[width=0.40\textwidth]{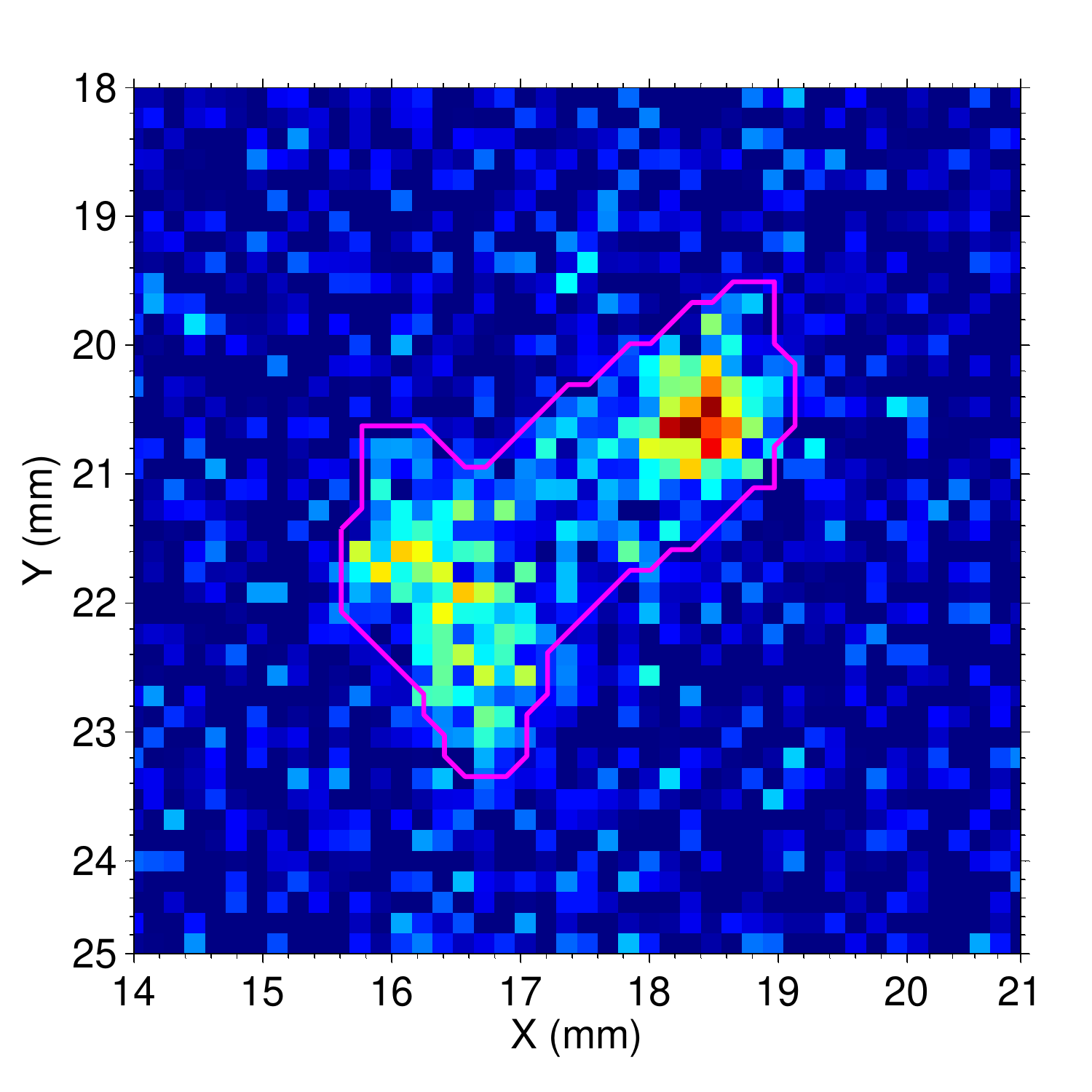}
	\label{fig:electronic3}}
	\qquad
	\subfloat[ ][$>$ 30 keVee electron recoils]{
	\includegraphics[width=0.40\textwidth]{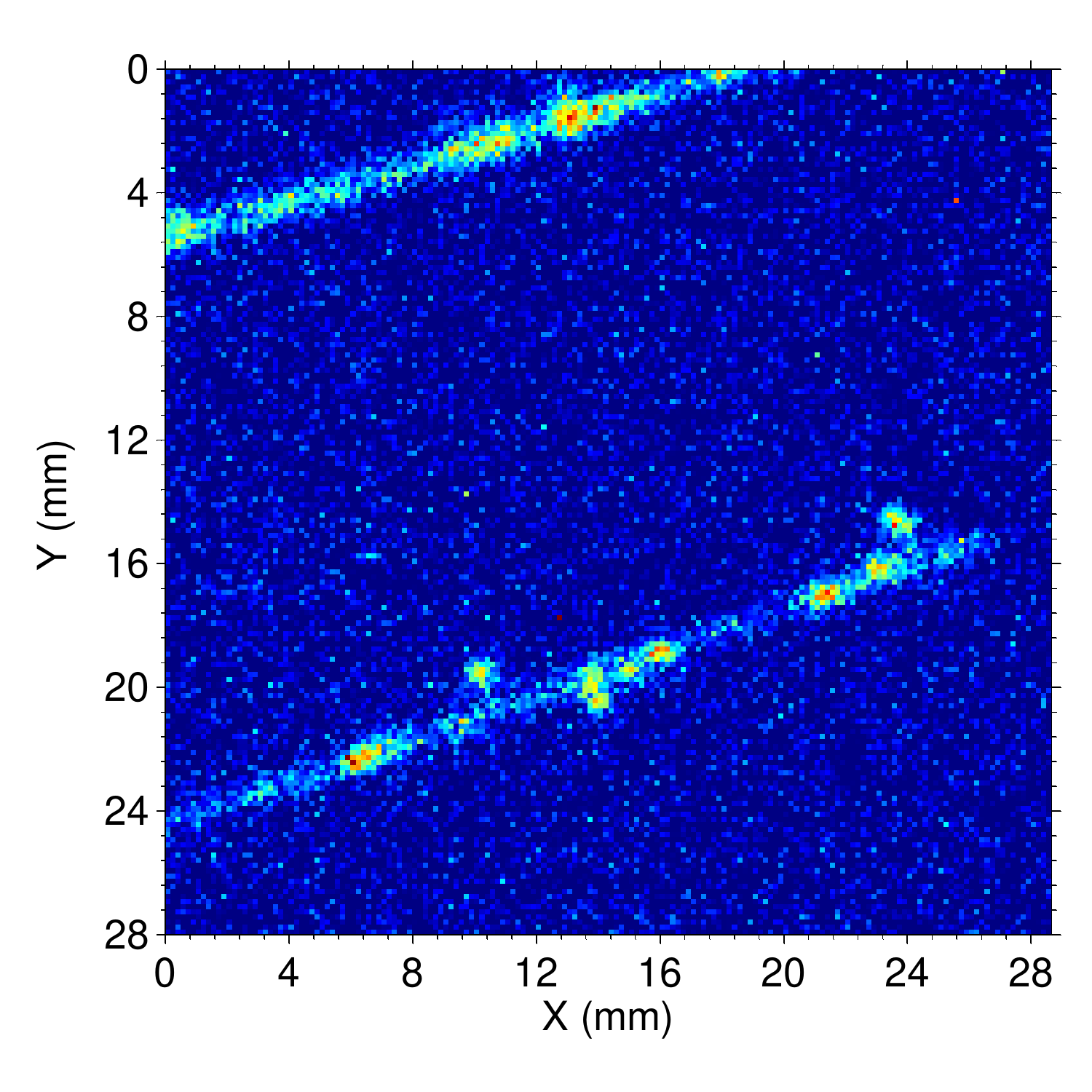}
	\label{fig:highenergyelectron}}	
	\caption[]{(a-d) Electronic recoils of different energies from the $^{252}$Cf and $^{60}$Co runs. The images have been contrast adjusted to enhance visualization.  The magenta contours trace out the track boundaries and are included as a visualization aid and to help illustrate the straggling of low energy recoils and the clumpy ionization deposition. (d) Two high energy electronic recoils containing smaller delta ray tracks emerging perpendicular to the primary electronic recoil track.  These image boundary crossing tracks were rejected from the analysis and were found only by visually scanning events by eye.}
	\label{fig:electronrecoilimages}
\end{figure*}

\begin{figure*}[]
	\centering
	\subfloat[Electron Recoil]{\includegraphics[width=0.40\textwidth]{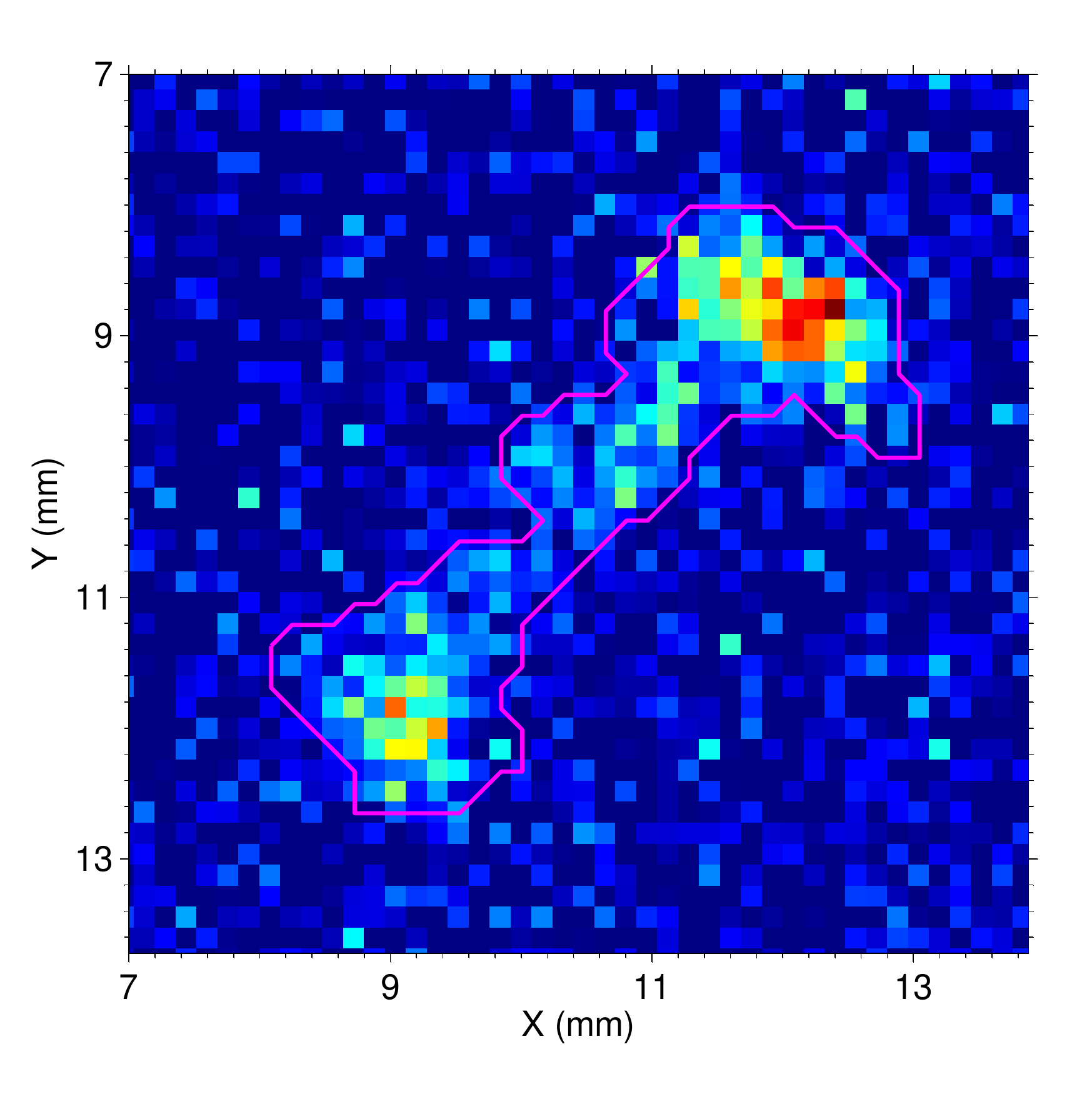}
	\label{fig:braggtrack}}
	\qquad
	\subfloat[Projected Bragg Curve]{\includegraphics[width=0.40\textwidth]{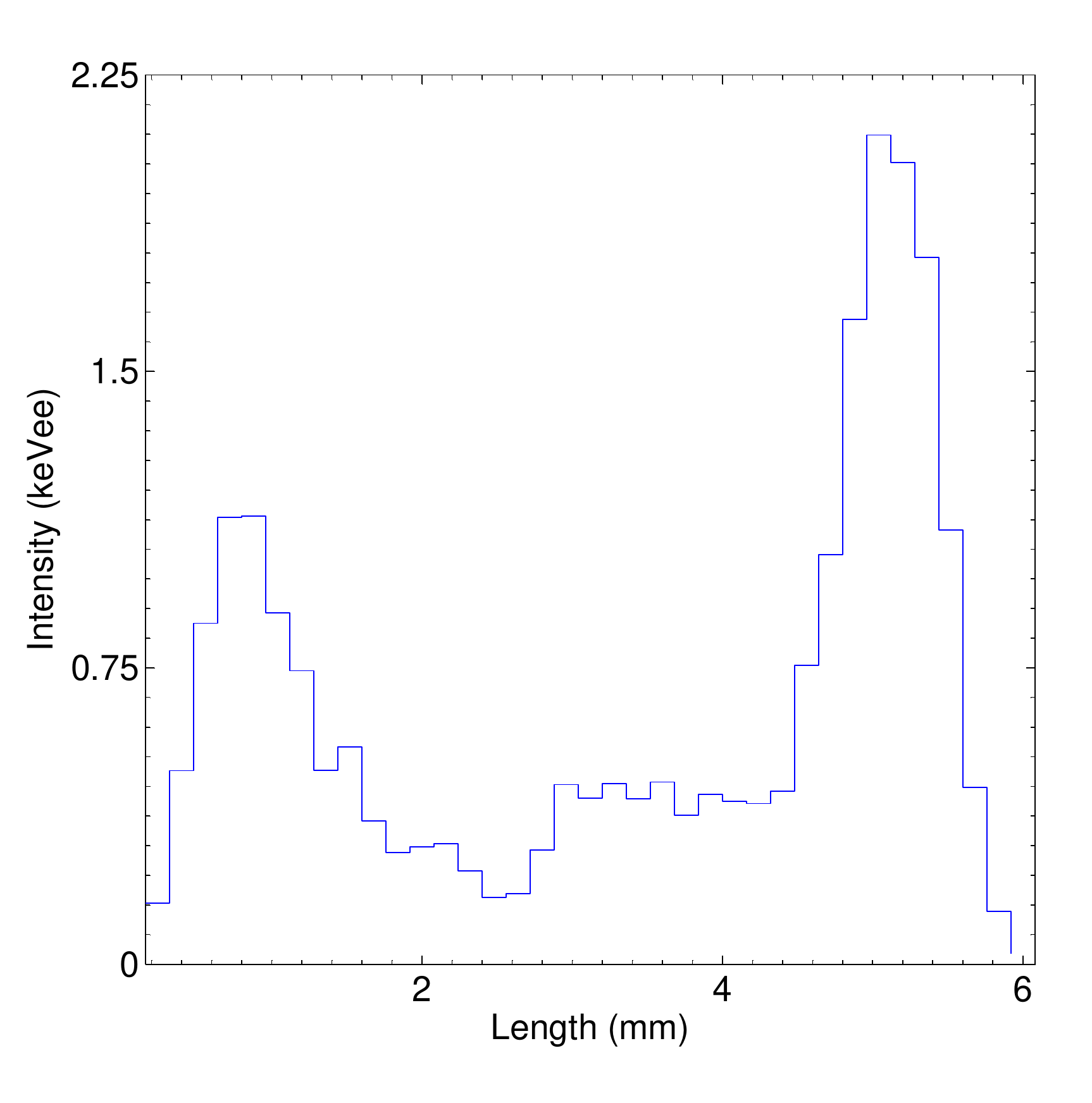}
	\label{fig:braggdist}}
	\caption[]{(a) An image of a 24 keVee electronic recoil in 100 Torr CF$_4$.  The magenta curve traces out the perimeter of the track and helps in visualizing the straggling of the recoil.  (b) The Bragg curve of the recoil, obtained from the projection of the track along the major axis of the fitted ellipse, shows the large energy fluctuations that are also clearly apparent in the CCD image.  A detector with a lower signal to noise ratio would only see the brightest region(s) of the track and possibly misidentifying them as a nuclear recoil(s).}
\label{fig:braggcurve}
\end{figure*}

\begin{figure*}[]
	\centering
	\subfloat[ ][28 keVr ($\sim 13$ keVee) nuclear recoil]{
	\includegraphics[width=0.40\textwidth]{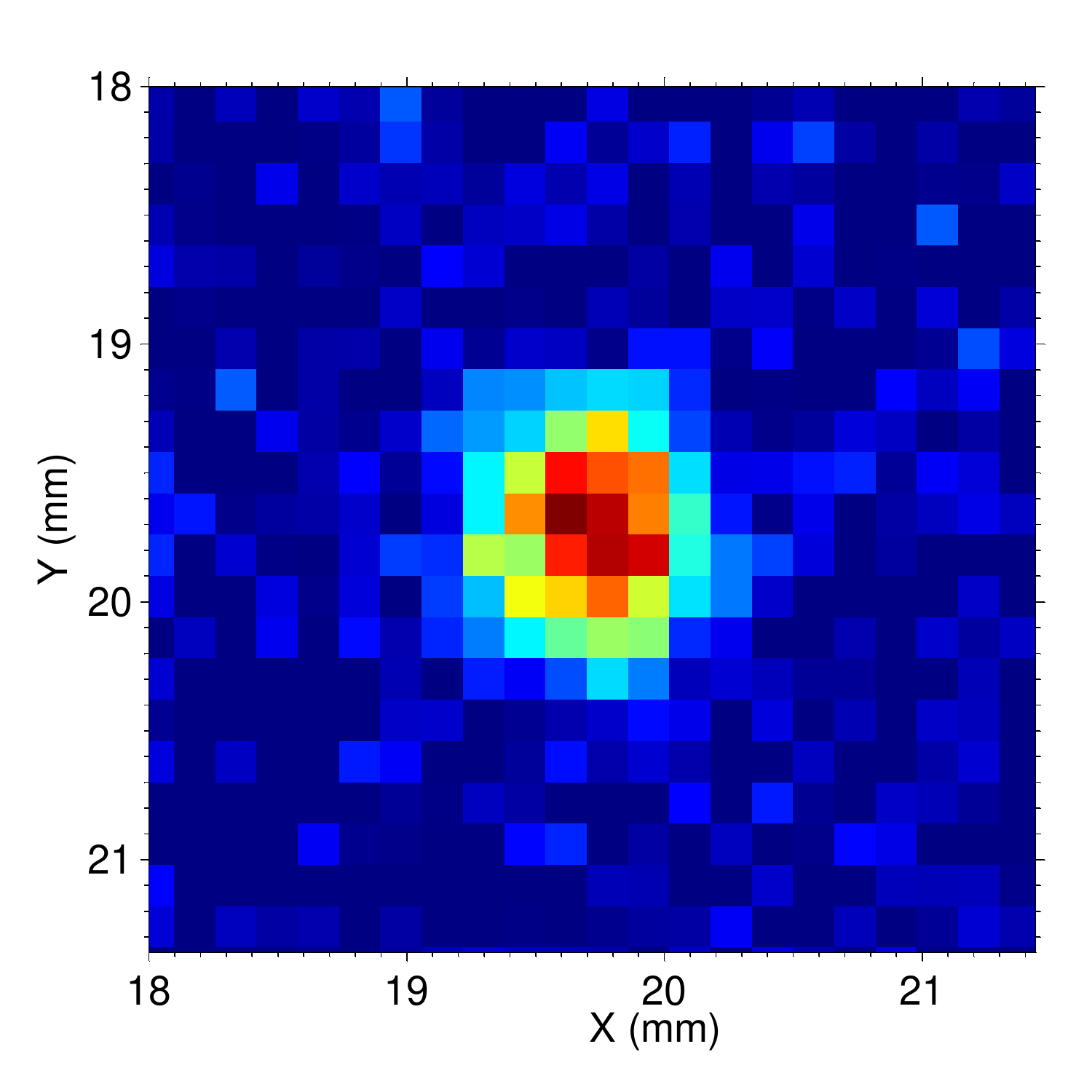}  
	\label{fig:nuclear1}}
	\qquad
	\subfloat[ ][53 keVr ($\sim 28$ keVee) nuclear recoil]{
	\includegraphics[width=0.40\textwidth]{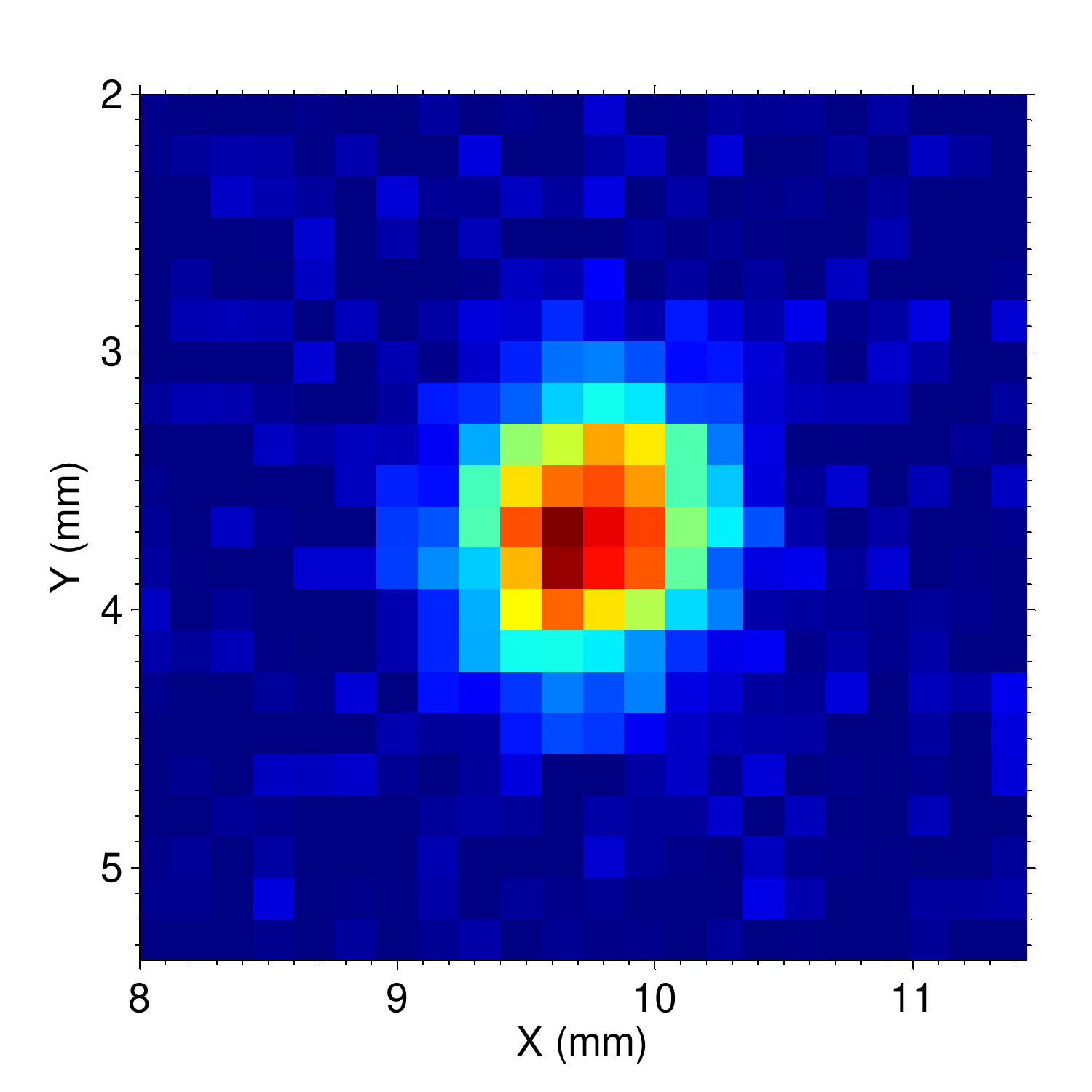}
	\label{fig:nuclear2}}	
	\qquad
	\subfloat[ ][104 keVr ($\sim 66$ keVee) nuclear recoil]{
	\includegraphics[width=0.40\textwidth]{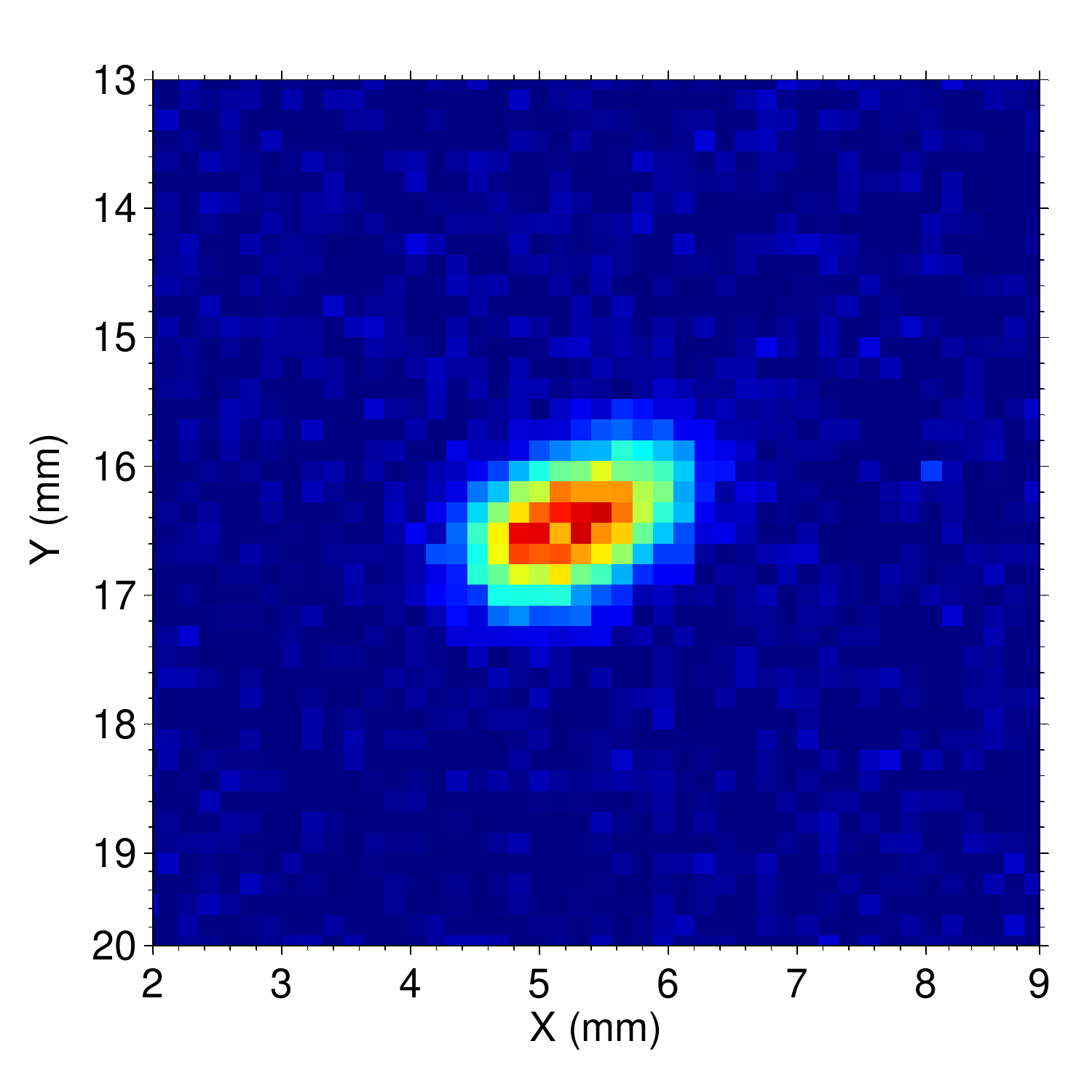}
	\label{fig:nuclear3}}
	\qquad
	\subfloat[ ][214 keVr ($\sim 160$ keVee) nuclear recoil]{
	\includegraphics[width=0.40\textwidth]{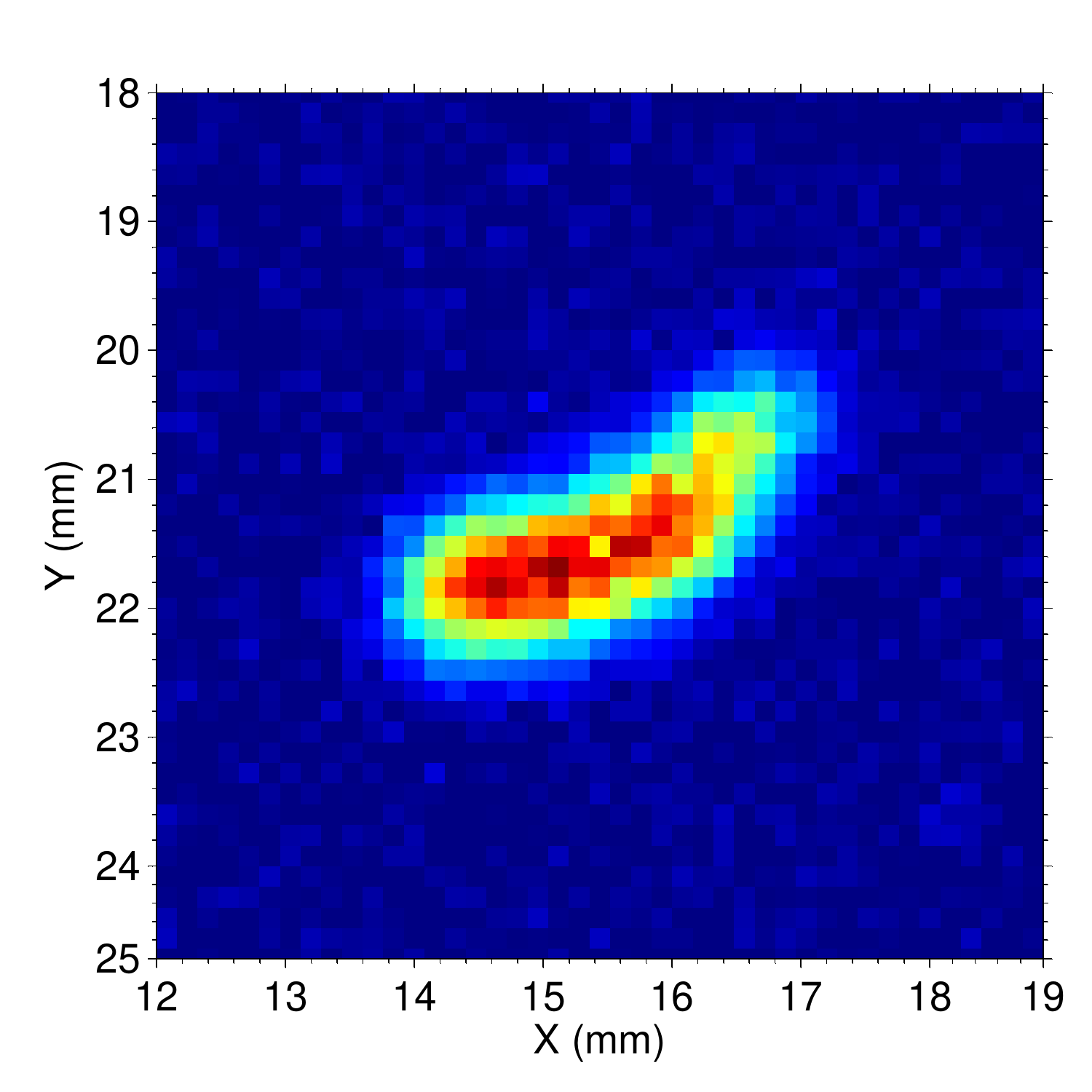}
	\label{fig:nuclear4}}	
	\caption[]{(a-d) Nuclear recoils from the $^{252}$Cf runs at various kinetic energies in 100 Torr CF$_4$ using the Hitachi quenching factors and assuming all the nuclear recoils are fluorine atoms.  The images have been contrast adjusted to enhance visualization and have different X/Y scales.  The directionality and asymmetry in the energy deposition often referred to as the head-tail signature become apparent for the highest three energy recoils.  In all images, the average neutron direction is from left to right.  Also note how the track in (a) deposits roughly the same amount of energy as the electronic recoils in Figures \ref{fig:electronic1} and \ref{fig:electronic2} but appear in a much tighter area with a higher intensity peak pixel value.  A lower signal-to-noise and lower resolution detector could fail to differentiate the track in (a) from that in Figure \ref{fig:electronic1}. }
	\label{fig:nuclearrecoilimages}
\end{figure*}

The resulting energy spectrum from the $^{252}$Cf run after applying the analysis cuts to remove CCD events and electron recoils is shown in Figure \ref{fig:spectrum-keV}.  Conversion from the measured energy in keVee to nuclear recoil energy in keVr is based on the fluorine quenching factors from Ref. \cite{hitachi}.  The spectrum rises from 20 keVr to 35 keVr, where it peaks, indicating that maximum nuclear recoil efficiency has been reached.  Thus, the effective discrimination threshold of this detector is approximately at 10 keVee ($\sim$23 keVr), see Figure \ref{fig:RPR_R2} and inset in Figure \ref{fig:recoilspectrum}.  At 100 Torr, this is the lowest discrimination threshold of any directional detector to date.  Nevertheless, our $\sim$10 keVee discrimination threshold is significantly above the detection threshold, which we estimate to be 2 keVee for a diffused, point-like event based on our $^{55}$Fe calibration data (see Figure \ref{fig:fe55spectrum}).  Our directional threshold is higher yet, $\sim$$2 \times$ the discrimination threshold, with the reason being that nuclear recoil tracks are shorter due to their higher $dE/dx$, and therefore become unresolved at higher energies; further details are discussed in a separate paper on directionality.  The importance of a low energy threshold, for both discrimination and directionality, is that it provides one path towards increasing the sensitivity of directional dark matter detectors.  In fact, it is critical for a low mass WIMP search as the recoil energy spectrum is shifted towards lower energies.  Finally, sample recoil images from the $^{60}$Co and $^{252}$Cf runs are shown in Figures \ref{fig:electronrecoilimages}, \ref{fig:braggcurve}, and \ref{fig:nuclearrecoilimages}.

\section{Discussion}
\label{Discussion}

The break down of discrimination in our detector below $\sim$10 keVee ($\sim$23 keVr) is due to a number of effects that lead to the convergence of electron and nuclear recoil tracks in the range versus energy parameter space.  These are due to physical effects, such as diffusion and energy-loss processes, as well as detector limitations.  
We discuss these below in Sections \ref{sec:discrimination threshold} and \ref{sec:detector improvements}, and describe possible ways to circumvent them to improve discrimination.

\subsection{The discrimination threshold}
\label{sec:discrimination threshold}

The discrimination at low energies is affected first and foremost by diffusion; as tracks fall below the resolution limit, range versus energy no longer works as a discriminant. Even if diffusion were suppressed, however, energy-loss processes affecting both electrons and nuclear recoils could pose fundamental limits to discrimination.  

For electrons the dominant effects are the well known energy-loss fluctuations and straggling (e.g., Figures \ref{fig:electronic2}, \ref{fig:electronic3}, and \ref{fig:braggcurve}), which give rise to a large spread in their range.  These effects becoming stronger at lower energies, pushing the short-track tail of the electron distribution below the diffusion limit.  There, these electron recoils merge with the nuclear recoil population (Figures \ref{fig:gamma12} and \ref{fig:neutron12}) which, with their much larger $dE/dx$, are already unresolved at $\sim$20 keVee ($\sim$40 keVr).  In addition, the probability for large angle scattering at low energies increases for electrons, producing a trajectory that is almost diffusive in nature.  As a result the energy is deposited into smaller unresolved regions of space, which, together with projection of the track to 2D, systematically biases the $dE/dx$ upward towards that of nuclear recoils.  In Figure \ref{fig:neutron12} events of a given energy affected in this manner have their $R_2$ underestimated and drop down into the nuclear recoil band.

For nuclear recoils the opposite trend occurs, whereby energy-loss in the ionization channel (the detectable $dE/dx$) decreases as the ions slow down, with other energy-loss channels making up the difference.  Both theoretical (Ref. \cite{hitachi}) and experimental (Ref. \cite{MIMAC-quench}) studies of the ratio of ionization to total energy-loss (the quenching factor) indicate values less than 0.25 for E $< 10$ keVr in a variety of gases and their mixtures.  This effect underestimates the energy of events of a given track length in Figure \ref{fig:neutron12}, pushing them leftward into the electronic recoil band. 

Thus, the detected $dE/dx$ for both classes of recoils converges at low energies, potentially posing a fundamental limit on discrimination using the range versus energy technique.  The energy where this occurs cannot be determined from our data where, as mentioned above, the limitations on discrimination are due to diffusion.  Progress toward this goal will require resolving tracks below our E $<10$ keVee threshold, for example by lowering the gas pressure to lengthen tracks.  From a practical perspective, directional gas TPCs have been shown to operate down to 20 Torr (\cite{Buckland94, Buckland97}), and with Thick GEMs (THGEMs) good gas gain has been demonstrated down to 0.5 Torr in certain gases \cite{shalemTHGEM1, shalemTHGEM2}.  So, measurements in the 10 - 40 Torr pressure range could feasibly map out the possible parameter space for discrimination below 10 keVee, and will likely lower the threshold as well.  Exploring gas mixtures with lower straggling and energy loss fluctuations should also be attempted.  All such efforts will be most critical for low mass WIMP searches where energy thresholds $<$ 10 keVr (note this is recoil energy, not ionization energy) are desired.  Depending on the degree of quenching, the detected energy in this regime could be as low as a few keVee, where achieving both discrimination and directionality could be extremely challenging.

\subsection{Detector improvements}
\label{sec:detector improvements}

\begin{figure*}[]
	\centering
	\subfloat[Noise Added]{\includegraphics[width=0.30\textwidth]{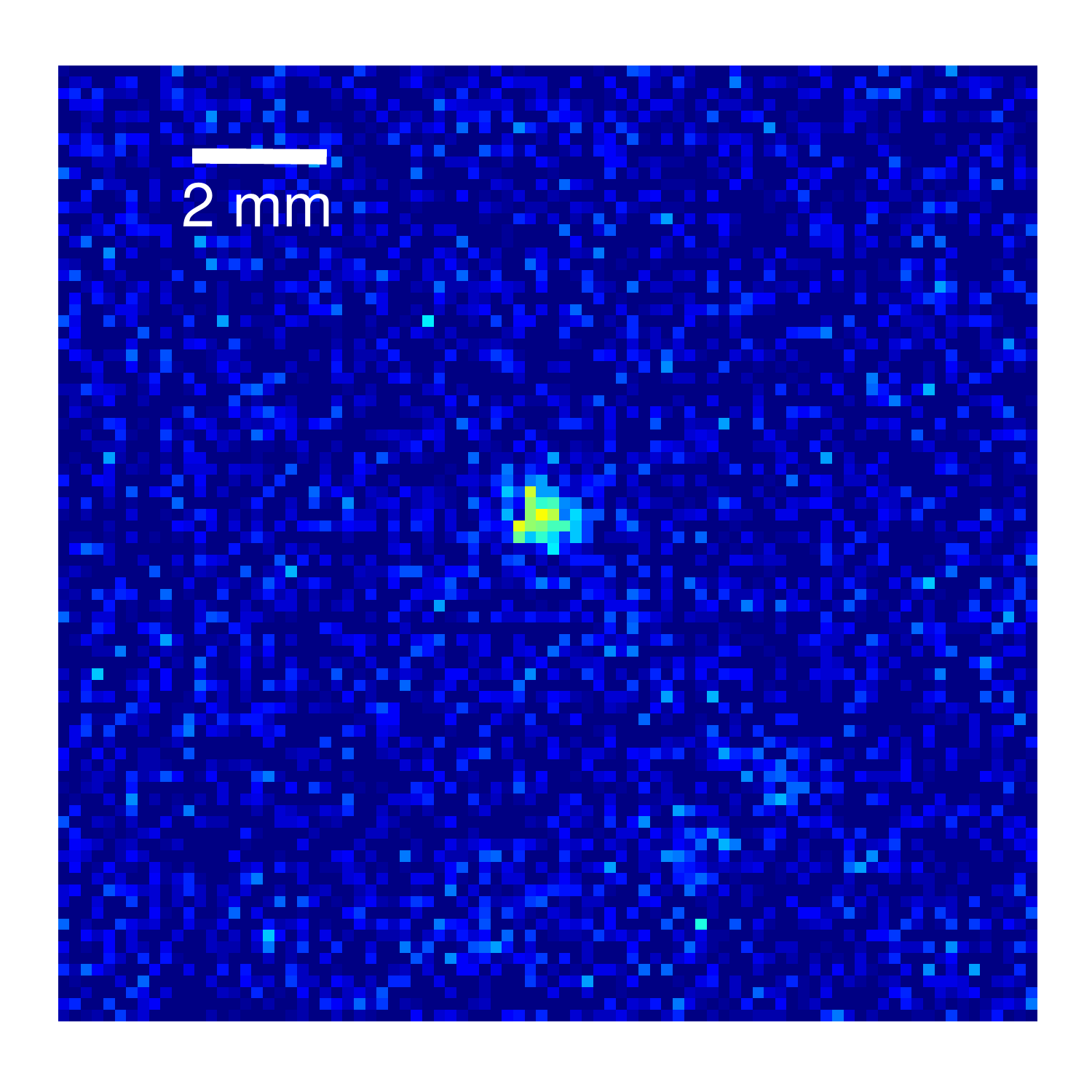}
	\label{fig:noiseadd}}
	\qquad
	\subfloat[Original]{\includegraphics[width=0.30\textwidth]{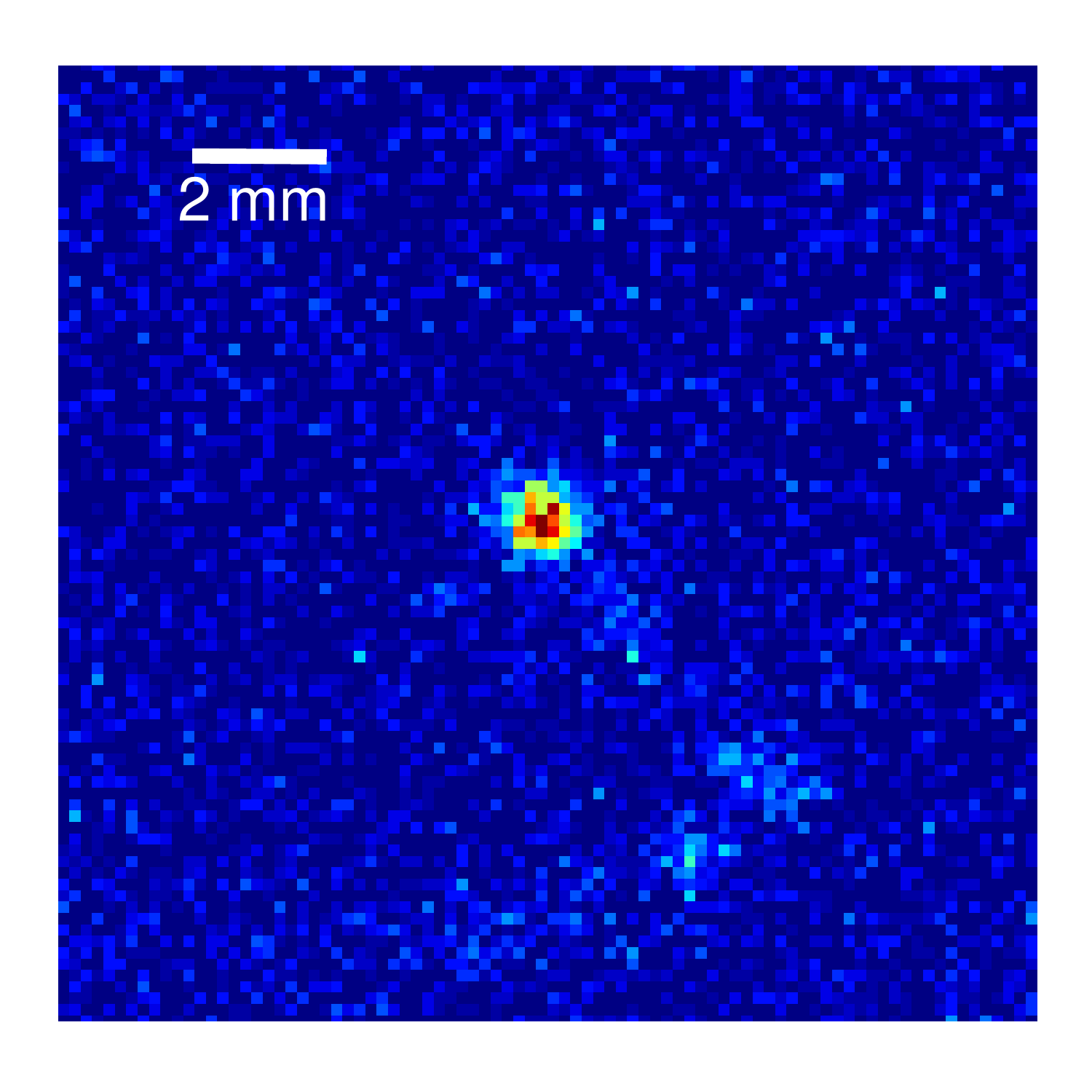}
	\label{fig:original}}
	\qquad
	\subfloat[Filtered Original]{\includegraphics[width=0.2975\textwidth]{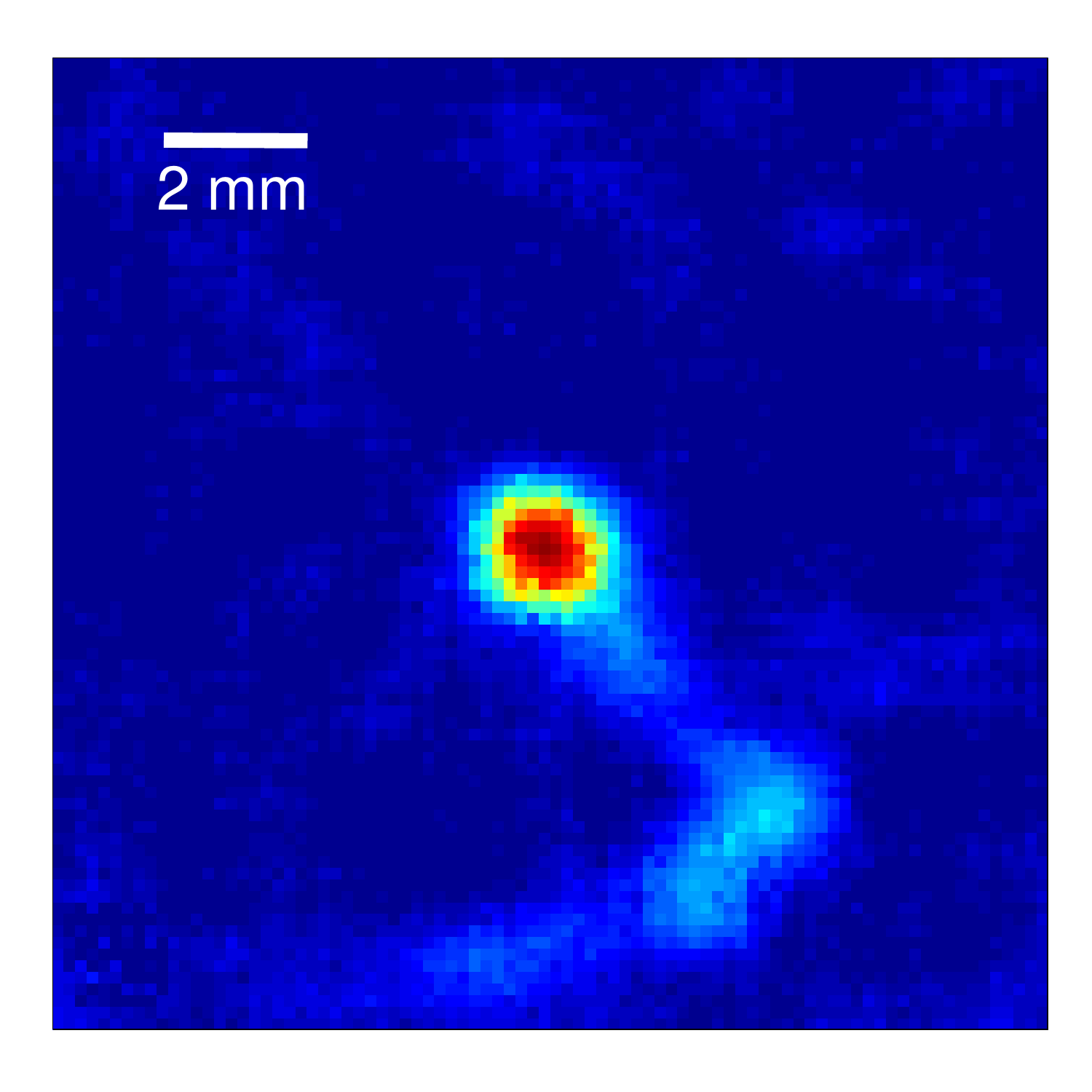}
	\label{fig:filteroriginal}}
	\caption[]{An $\sim$16 keVee event from the $^{252}$Cf run shown with different signal-to-noise levels.  The middle panel (b) shows the original calibrated CCD image, and the left panel (a) shows the same but with 50\% higher noise added in software.  The right panel (c) is the original image with a Gaussian noise reduction filter applied.  See text for details. }
	\label{fig:electronSNR}
\end{figure*}

Besides lowering the pressure and optimizing gas mixtures, improvements in the detector itself could also lead to better discrimination and directionality.  The three detector parameters that we believe play a critical role for this are signal-to-noise, resolution, and tracking dimensionality (discussed in  Section \ref{sec:Background Discrimination 1D 2D 3D}).  We restrict our discussion to an optical detector of the type used here, but many of the ideas apply to charge readout detectors as well.  

A benefit of signal-to-noise, especially where discrimination is concerned, is that it enables full mapping of electron tracks that have large energy-loss fluctuations of the type seen in Figures \ref{fig:electronic2}, \ref{fig:electronic3}, and \ref{fig:braggcurve}.  These tracks show regions with high energy-loss interspersed with barely discernible regions of low energy-loss.  A detector with low signal-to-noise would detect only the peak regions, which, due to their systematically higher $dE/dx$, would look like nuclear recoil tracks.  This is illustrated in Figure \ref{fig:electronSNR}, where we show three images of an event from the $^{252}$Cf run, each with a different signal-to-noise level.  Figure \ref{fig:original} is the original calibrated CCD image obtained by our detector, showing a track with a high density segment and a suggestion of a very faint tail.  In Figure \ref{fig:noiseadd}, we artificially added about 50\% more noise to the image in software.  In this image the long faint tail of the recoil is lost in the noise, leaving just the bright high density region that could easily be misidentified as a low energy nuclear recoil track.  Finally, in Figure \ref{fig:filteroriginal} the original image has been processed using a Gaussian noise reduction filter.  The long faint tail is now clearly visible as is its connection to the brighter segment, leaving little doubt that this is an electron track.

The two obvious paths to achieving high signal-to-noise are to lower the noise and/or increase the signal.  The former approach would require a reduction in the CCD camera system noise, which is usually dominated by the read noise for short exposures, but the tradeoff is slower readout speed or more costly multi-node readouts.  In the direction of increasing signal, there are many approaches that could be taken.  The first is boosting the CCD sensor quantum efficiency, which, for the back-illuminated CCD used here, is already highly optimized.  Secondly, one can increase signal through better light collection with a more efficient optical system (faster lens), and/or a setup that allows for more light collection by decreasing the distance between the GEM and lens.  However, the latter approach requires a sacrifice of imaging area whereas the former requires a potentially uneconomical and sophisticated custom lens design.  Although both are potential drawbacks for scale-up to large detector volumes, these approaches should be considered if cheaper CCDs or other technologies become available in the future.

Another approach is to increase the light output by selecting gas mixtures with higher avalanche photon yield, defined as the number of photons per secondary electron released during amplification, or by increasing the absolute gas gain.  Although we have achieved very high gas gain in our detector with the triple-GEM stack, the light yield of CF$_4$, albeit one of the better scintillating gases, could be improved further.  For example, the addition of Ar at high concentrations has been shown to increase the photon yield of pure CF$_4$ \cite{kaboth, pansky} from $\sim$0.3 to $\sim$0.7 \cite{fragalightyield}.  Although a number of excellent gas scintillators exist, consideration of the spectrum (e.g., optical vs. UV) and whether the target is optimal for the specific WIMP search (e.g., spin-dependent vs. spin independent) must be taken into account.  The gas gain could also be increased, but saturation effects have been noted at high gains \cite{kozlov}, where both the gain and photon yield are charge density dependent.  This could have a deleterious effect on both energy and directional sense determination.

Detector resolution, also critical for both directionality and discrimination, is governed by various design and operation choices such as the readout pitch, gas mixture, pressure and diffusion.  How pressure can be used to vary track lengths and how the choice of gas can effect fluctuations and straggling were briefly discussed above, so we focus on the other two factors here.  Diffusion can be reduced by limiting the maximum drift distance and by making a judicious choice of gas mixture. 
Although CF$_4$ exhibits relatively good diffusive characteristics for an electron drift gas, negative ion gases such as CS$_2$, which drift in the thermal regime, provide the lowest diffusion possible without employing magnetic fields (see Section \ref{Directional Detection Challenges}).  The low diffusion in our small detector, $\sigma\sim$ 0.35 mm, is not far from the average value of $\sigma\sim$ 0.5 mm achieved with CS$_2$ over a 50 cm drift in the DRIFT detector.  Thus, the diffusion achieved in our detector is a reasonable goal for a large scale directional experiment, and any meaningful reduction would likely require other techniques.

In principle, the readout pitch of the detector should be fine enough to extract the maximum information possible with the given diffusion.  The effective pitch in our detector, which is due to a combination of the GEM pitch, 140 \si{\um}, and the CCD binned pixels, 165 \si{\um}, was a little less than half the sigma due to diffusion, $\sigma \sim 0.35$ mm.  This allowed tracks to be measured with a sufficient number of independent samples to extract features, such as energy-loss fluctuations and asymmetry in ionization, important for discrimination and directionality. 

A good example demonstrating this for discrimination is found by comparing the images of the electronic recoil and nuclear recoil shown in Figures \ref{fig:electronic1} and \ref{fig:nuclear1}, respectively.  Both have similar detected energy but the electronic recoil looks more disperse with larger fluctuations, and the nuclear recoil more concentrated and smoothly distributed.  These differences are consistent with the energy-loss processes discussed above, and could be used in more sophisticated algorithms to improve discrimination and directionality.  The finer pitch also opens the door to deconvolution techniques, such as those used in astronomy (for example, see \cite{MCS}), which could be applied to achieve better resolution.

\subsection{Background Discrimination: 1D, 2D, and 3D}
\label{sec:Background Discrimination 1D 2D 3D}

\begin{figure*}[]
	\centering
	\subfloat[X Component]{\includegraphics[width=0.30\textwidth]{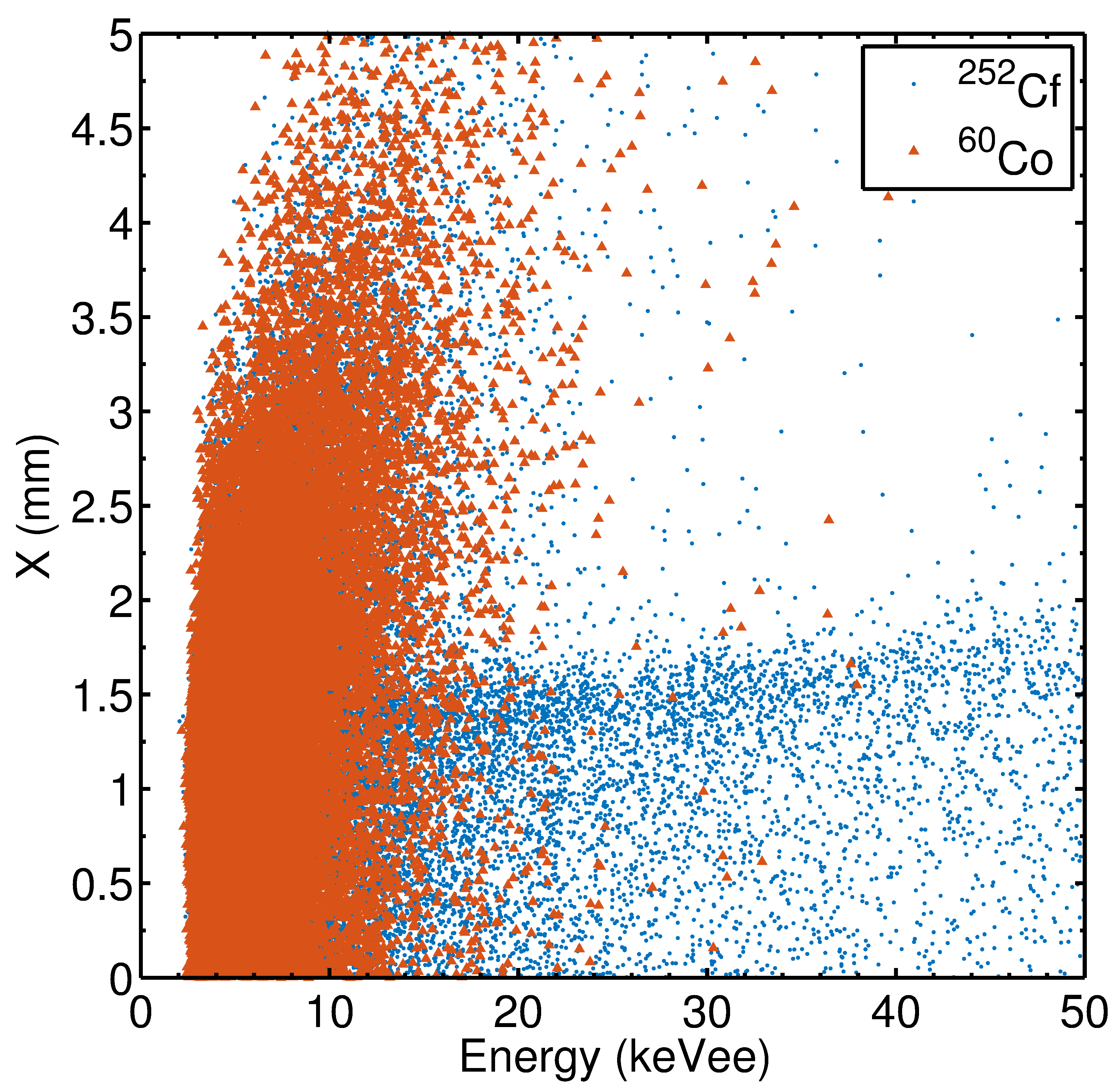}
	\label{fig:XmmOverlay}}
	\qquad
	\subfloat[Y Component]{\includegraphics[width=0.30\textwidth]{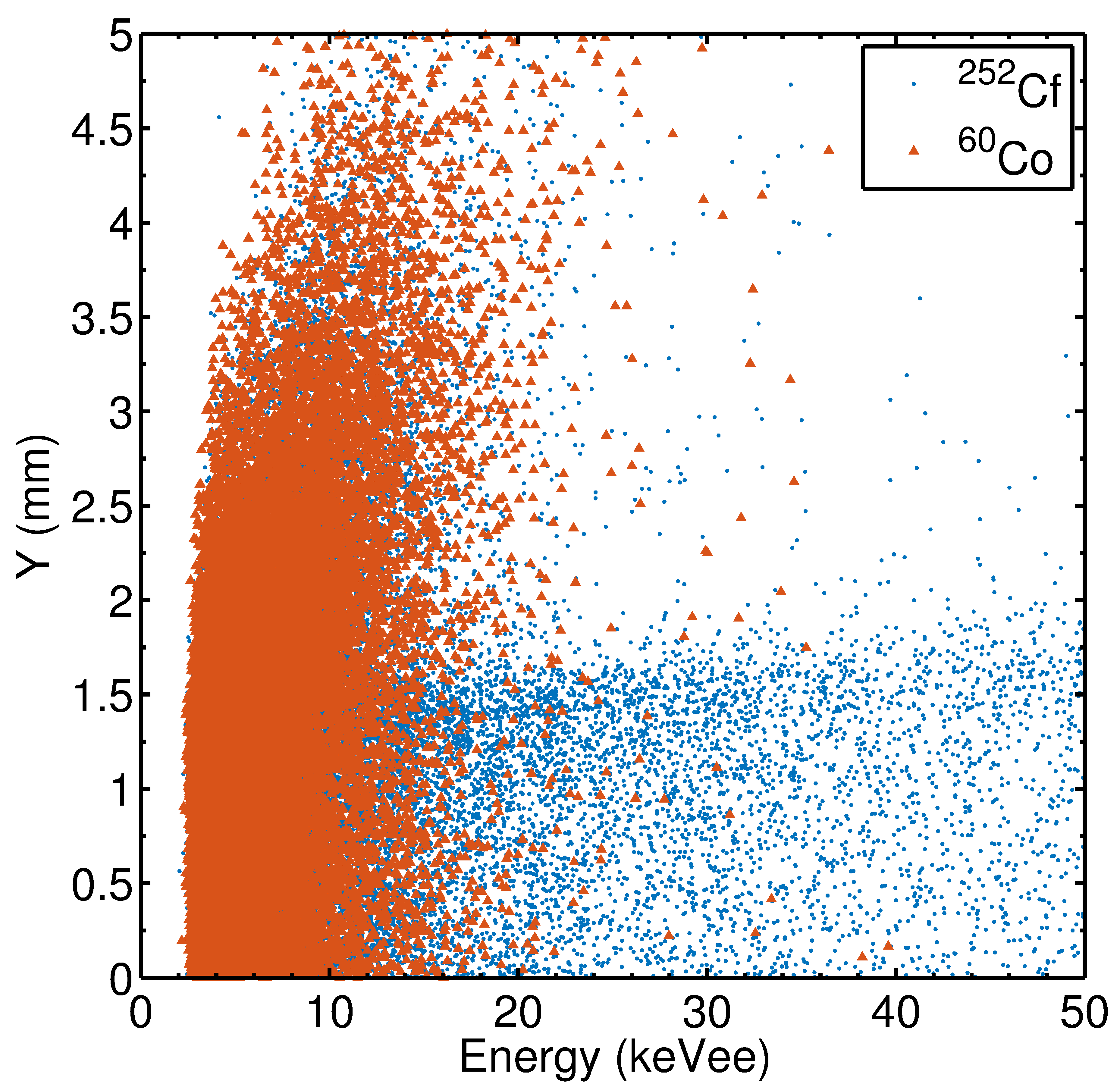}
	\label{fig:YmmOverlay}}
	\qquad
	\subfloat[$R_2$ Length]{\includegraphics[width=0.30\textwidth]{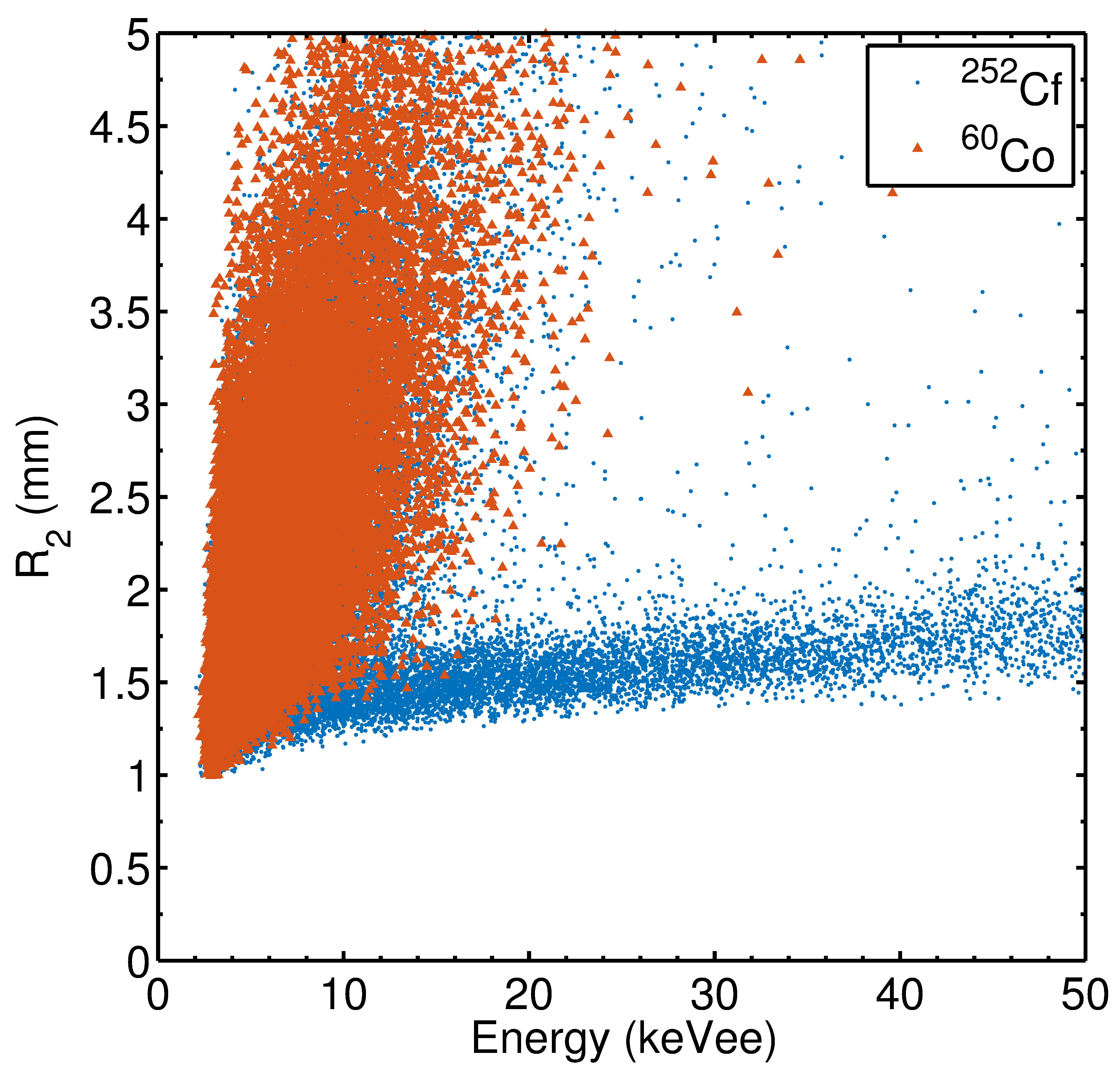}
	\label{fig:R2mmOverlay}}
	\caption[]{(a) and (b)  Plots of the X and Y, 1D components of the track length vs. energy for the $^{60}$Co run overlaid on top of the $^{252}$Cf run. These show the approximate level of discrimination that a 1D detector would achieve.  (c) The 2D projected track length vs. energy data from the same two data runs reproduced for comparison (from Figures \ref{fig:gamma12} and \ref{fig:neutron12}).  For all 3 panels the RPRs from the $^{60}$Co have been removed to show the true separation between the electronic recoil and the nuclear recoil bands.  In the 2D data there is separation of the bands down to about 10 keVee (23 keVr), but in the 1D data (a-b), events from the electronic recoil bands are leaking into the nuclear recoil region up to energies $>$ 35 keVee (62 keVr).}
	\label{fig:2D1D}
\end{figure*}

The detector parameter that is arguably most critical for good discrimination and directionality is the number of independently measured track components.  Although full 3D track reconstruction is preferred, any benefit it brings to discrimination or the directional sensitivity must be justified relative to the cost increase or added design and operational complexity.  Here we study the improvement in discrimination from 1D to 2D to 3D and, except for brief remarks below, postpone the discussion on directionality for a separate paper.  

We begin by studying the difference in discrimination power between 1D and 2D.  For this we took our 2D data from the $^{60}$Co and $^{252}$Cf runs and reduced them to 1D.  We have defined the X(Y) component of the track length as $R_{2}\cos\theta$($R_{2}\sin\theta$), where $\theta$ is the reconstructed angle of the track in the X-Y plane.  Of course, this artificially extends the 1D track length down to zero, whereas the diffusion in a real 1D detector would impose a minimum.  Nevertheless, this effect, which is apparent in Figures \ref{fig:XmmOverlay} and \ref{fig:YmmOverlay}, does not change the relative comparison we wish to make here.  In addition, to better gauge the separation between the electronic and nuclear recoil bands, the 65 events associated with RPRs in the $^{60}$Co were removed from this dataset.  

In Figures \ref{fig:XmmOverlay} and \ref{fig:YmmOverlay} the X and Y, 1D components of the tracks are plotted as a function of energy, respectively.  As one would expect, the electronic recoil bands in these two figures are very similar because their recoil directions are distributed more or less isotropically in the imaging plane.  The nuclear recoil band, although smeared out greatly in both figures, is marginally denser in Figure \ref{fig:XmmOverlay} because the neutrons were directed along the X-axis. For comparison, an overlay of the 2D, $R_{2}$ versus $E$ data from both runs is also shown in Figure \ref{fig:R2mmOverlay}. 

\begin{figure*}[]
	\centering
	\subfloat[2D Reconstruction]{\includegraphics[width=0.45\textwidth]{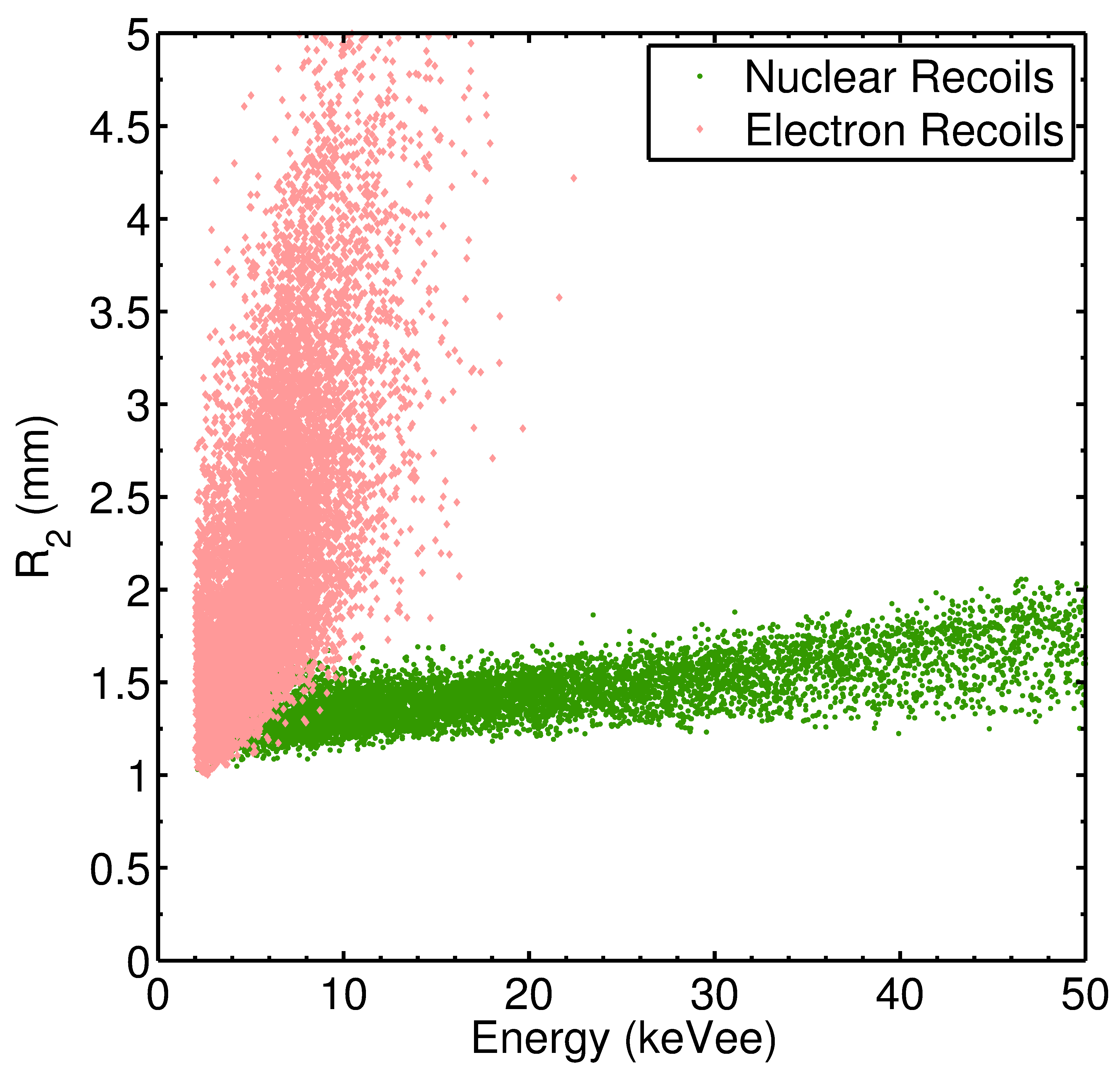}
	\label{fig:sim2D}}
	\qquad
	\subfloat[3D Reconstruction]{\includegraphics[width=0.45\textwidth]{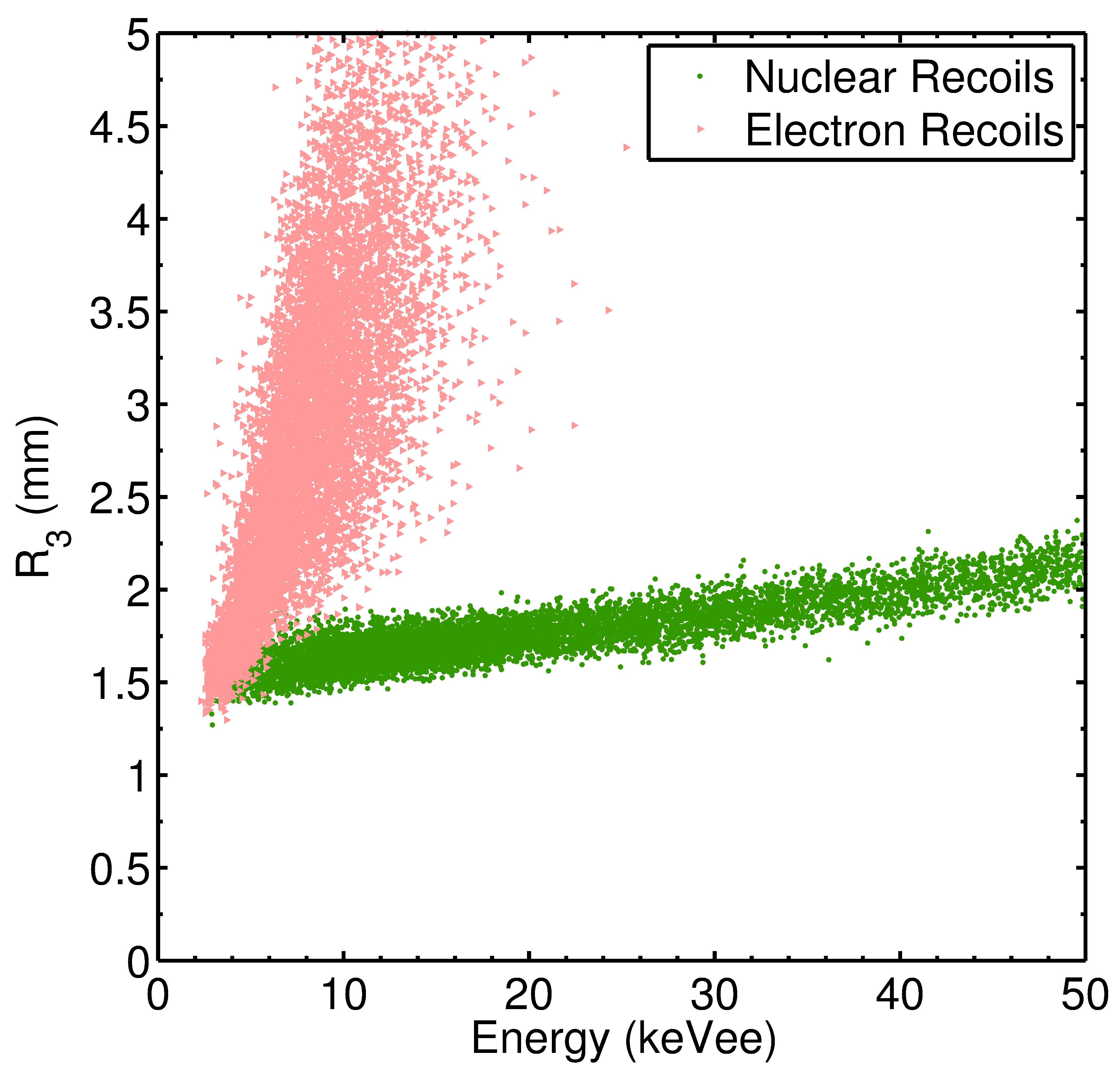}
	\label{fig:sim3D}}
	\caption[]{Simulation of range vs. energy for fluorine and electron recoils in 100 Torr CF$_4$ for 2D (a) and 3D (b) track reconstructions.  In the 2D reconstruction (a), events from the electron band leak into the nuclear band up to energies of $\sim$9 keVee.  But in the 3D reconstruction (b) events from the two bands are separable down to energies of $\sim$6 keVee. }
	\label{fig:globalsim}
\end{figure*}

The results show significantly better discrimination with 2D tracks versus 1D.  In 2D, discrimination is achieved down to $\sim$10 keVee (23 keVr), whereas in 1D, electron events from the $^{60}$Co run have strayed into the nuclear recoil band out to energies of $\sim$35-40 keVee ($\sim$65 keVr).  This effectively puts the discrimination energy threshold of the 2D data at a factor $\sim$3 lower than the 1D data, which would correspond to a factor of $\sim$7(70) times higher detection sensitivity for a WIMP of mass 100(30) GeV$\cdot$c$^{-2}$ scattering off fluorine through spin-independent interaction.  Perhaps cuts made on other track parameters could be used to reduce this gap, but it is unlikely that 1D discrimination would improve to extend the threshold much below 30 keVee (55 keVr).   
 
Next, with the aid of simulations we explored the potential difference in discrimination capability between a 2D vs. 3D detector.  The simulation program SRIM(CASINO) \cite{ziegler}(\cite{casino}) was used to simulate nuclear(electronic) recoil tracks with an isotropic distribution in 3D and with the same energy distribution as our calibration data.\footnote{We note that the nuclear recoil simulations do not take into account secondary recoils.}  Each simulated track was then projected onto the image plane with the pixellization, noise, and signal adjusted to match those of our CCD detector.  The diffused and signal-to-noise adjusted projected track from each image plane (XY, XZ, YZ) was analyzed using the same image analysis algorithm as the one used for our neutron and gamma data (Section \ref{sec:Co-60 and Cf-252 Runs}).  

The results of these simulations are shown in Figures \ref{fig:sim2D} and \ref{fig:sim3D}.  As expected, the nuclear recoil band for the case of 3D reconstruction (Figure \ref{fig:sim3D}) is much tighter than the 2D case (Figure \ref{fig:sim2D}).  The 3D distribution of the electron band is still dominated by the effects of straggling and energy loss fluctuations (discussed in Section \ref{sec:discrimination threshold}), resulting in large scatter in both range and energy.  However, in the region where the two bands intersect, the 3D electron events are more tightly distributed than in 2D, yielding better separation from the nuclear recoils.  This results in about a $\sim$35\% lower discrimination threshold, which, not surprisingly, is not as large as the difference seen between the 1D and 2D data.  Nevertheless, when combined with lower, $\sim$10-20 Torr, operating pressures and better track reconstruction algorithms, 3D could push the discrimination threshold into the few keVee region of interest for low mass WIMP searches.

With regards to the WIMP directional signature, the advantage of 3D track reconstruction in directional searches has been a subject of numerous studies and discussions.  Using the criteria of number of events needed to reject isotropy, these studies show only a factor few difference between 2D and 3D when perfect HT sense recognition is assumed \cite{MorganGreen20053D, MorganGreen20052D, Copi2007}.  If other variables are included then even 1D appears competitive \cite{Billard2015}. This would seem to suggest that multi-dimensional tracking is something desired but not absolutely necessary.  There are two caveats to this, however.  The first is that the assumption of perfect HT sense recognition is unrealistic, and we argue that higher dimensionality is needed even to approach this goal.  The second is that, in the case of low pressure TPCs, discrimination power and directional sensitivity are coupled to the tracking dimensionality of the detector.  As the primary discriminant is the stopping power, $dE/dx$, robust discrimination requires high quality measurements of both energy and track range.  The latter, as we have shown here, is best accomplished with a 3D detector.  A more extensive discussion on the relationship between tracking dimensionality and directional sensitivity is reserved for a separate paper.

\section{Conclusion and Prospects}
\label{Conclusion and Prospects}

In this work we have described a small high resolution, high signal-to-noise GEM-based TPC with a 2D CCD readout.  The detector was designed to make detailed studies of low energy electron and nuclear recoil tracks for the purpose of directional dark matter searches.  Detector performance was characterized using alpha particles from $^{210}$Po, X-rays from $^{55}$Fe, gamma-rays from $^{60}$Co, and $\sim$MeV neutrons from $^{252}$Cf.  Stable gas gains upward of $10^5$ were achieved in 100 Torr of pure CF$_{4}$ with a triple-GEM cascade, resulting in a very high signal-to-noise.  This, together with an effective 165 \si{\um} track sampling and low diffusion, $\sigma \sim$ 0.35 mm, provided the means for detecting events with energies down to a few keVee.  

With our $^{60}$Co and $^{252}$Cf data we also studied discrimination between electronic and nuclear recoils.  Using the standard range versus energy technique, relatively simple selection criteria were used to demonstrate excellent discrimination down to $\sim$10 keVee, or $\sim$23 keVr recoil energy.  This result, the best to date at 100 Torr, was especially aided by the high spatial resolution and signal-to-noise of the detector.  Without the latter, the large energy-loss fluctuations suffered by low energy electrons would cause only the peak intensity regions of the tracks to be detected.  Such tracks would be reconstructed with their $dE/dx$ and track lengths systematically too high and too low, respectively, resulting in these events being misidentified as nuclear recoils.  That both high spatial resolution and high signal-to-noise are necessary for good discrimination is an important result of this work.

Pushing the discrimination threshold to even lower energies is an important future goal, especially critical for directional low mass WIMP searches.  Our $\sim$10 keVee threshold is due to a combination of diffusion and electron straggling, which results in the merger of the electron and nuclear recoil populations in the $R_2$ versus $E$ parameter space.  Two paths around this are to lower the gas pressure to lengthen tracks, and full 3D track reconstruction.  The former would allow mapping the parameter space for discrimination below our 10 keVee threshold and find any fundamental limit if it exists.  The influence of track dimensionality on discrimination was also investigated and, using our data, we found that the 1D threshold is a factor $\sim$3 higher than 2D.  Using simulations we also found better separation between the electron and nuclear recoil populations in 3D versus 2D, resulting in about $\sim$35\% lower discrimination threshold in 3D.  In addition to these two strategies, better analysis techniques that take full advantage of the difference seen between electron and nuclear recoil tracks (e.g., compare Figures \ref{fig:electronic1} and \ref{fig:nuclear1}) should be investigated in the future.

Finally, the data obtained in this work can also be used to characterize the directionality of the nuclear recoil tracks.  With their higher $dE/dx$, we find that these tracks become unresolved at energies around $\sim$20 keVee ($\sim$40 keVr), resulting in a directionality threshold that is a factor $\sim$2 higher than the discrimination threshold.  That these two thresholds are not the same is another important result from this work, which, given the differences in the energy-loss processes of electrons and nuclear recoils, is perhaps not surprising.  A detailed analysis of the directional signature, and its implications for WIMP detection, will be described in a separate paper.   

\section*{ Acknowledgements}
\noindent This material is based upon work supported by the NSF under Grant Nos. 0548208, 1103420, and 1407773.




\end{document}